\newif\ifsubmode
\submodefalse

\newif\ifprintfig
\printfigtrue

\newif\ifemulate
\emulatefalse

\ifsubmode
  \documentstyle[12pt,aasms4,epsf]{article}
  \received{}
  \accepted{}
  \journalid{}{}
  \articleid{}{}
\else
   \ifemulate
     \documentstyle[emulateapj,epsf]{article}
     \submitted{{\it Accepted for publication in AJ}}
   \else
     \documentstyle[11pt,aaspp4,epsf]{article} 
     \slugcomment{{\it Accepted for publication in AJ}}
   \fi
\fi

\lefthead{Rest et al.}
\righthead{WFPC2 images of the central regions of early-type galaxies}

\newcommand{\etal}{{et al.~}}

\newcommand{\lta}{\lesssim}
\newcommand{\gta}{\gtrsim}

\newcommand{\kmsmpc}{\>{\rm km}\,{\rm s}^{-1}\,{\rm Mpc}^{-1}}
\newcommand{\kms}{\>{\rm km}\,{\rm s}^{-1}}

\newcommand{\Msun}{\>{\rm M_{\odot}}}

\newcommand{\submodefigcaption}[2]{
\ifsubmode
\figcaption{#2}
\else
\printfigtrue
\addtocounter{figure}{1}
\newcommand{#1}{#2}
\fi
}

\newcommand{\dofig}[4]{
\ifprintfig
\clearpage
\begin{figure}
\epsfxsize=#3 truecm
\epsfysize=#4 truecm
\centerline{\epsfbox{#2}}
\ifsubmode
\vskip3.0truecm
\addtocounter{figure}{1}
\centerline{Figure~\thefigure}
\else\figcaption{#1}\fi
\end{figure}
\fi
}

\begin{document}

\title{WFPC2 Images of the Central Regions of Early-Type Galaxies --
       I. The Data}

\author{Armin Rest, Frank C. van den Bosch\altaffilmark{1,2}}
\affil{Department of Astronomy, University of Washington, Seattle, 
       WA 98195, USA}

\author{Walter Jaffe}
\affil{Leiden Observatory, P.O. Box 9513, 2300 RA Leiden, The Netherlands}

\author{Hien Tran, Zlatan Tsvetanov, Holland C. Ford, James Davies, 
        Joanna Schafer}
\affil{Department of Physics \& Astronomy, Johns Hopkins University,
       Baltimore, MD 21218, USA}


\altaffiltext{1}{Hubble Fellow}
\altaffiltext{2}{Current Address: Max-Planck Institut f\"ur
Astrophysik,   Karl-Schwarzschild   Strasse  1, Postfach   1317, 85741
Garching, Germany}


\ifsubmode\else
  \ifemulate\else
     \clearpage
  \fi
\fi


\ifsubmode\else
  \ifemulate\else
     \baselineskip=14pt
  \fi
\fi


\begin{abstract}
  We present high resolution $R$-band images of the central regions of
  67 early-type galaxies obtained   with the Wide Field and  Planetary
  Camera  2  (WFPC2) aboard  the Hubble Space   Telescope (HST).  This
  homogeneously selected  sample    roughly  doubles the   number   of
  early-type galaxies that have now been imaged at HST resolution, and
  complements similar  data on the central   regions of radio galaxies
  and the  bulges of spiral  galaxies.  Our sample strikingly confirms
  the  complex  morphologies of  the   central  regions of  early-type
  galaxies which have become  apparent from previous studies with HST.
  In particular, we  detect dust, either  in the form of nuclear disks
  or with  a filamentary distribution, in  43 percent of all galaxies,
  in good  agreement  with previous  estimates.  In addition,  we find
  evidence for  embedded stellar disks  in a remarkably large fraction
  of 51 percent.  In 14 of those galaxies the disk-like structures are
  misaligned with the main galaxy,  suggesting that they correspond to
  stellar bars in S0 galaxies.  We analyze  the luminosity profiles of
  the galaxies in our sample, and classify galaxies according to their
  central cusp  slope.  To a large  extent we confirm the results from
  previous  HST  surveys in  that  early-type galaxies  reveal a clear
  dichotomy: the  bright ellipticals ($M_B  \lta -20.5$) are generally
  boxy   and have  luminosity  profiles  that break  from steep  outer
  power-laws to shallow inner cusps (referred  to as `core' galaxies).
  The  fainter  ellipticals, on the  other  hand, typically have disky
  isophotes and luminosity profiles that lack a clear break and have a
  steep  central cusp (referred  to   as `power-law' galaxies).    The
  advantages and shortcomings  of classification schemes utilizing the
  extrapolated central cusp slope $\gamma$   are discussed, and it  is
  shown  that  $\gamma$  might   be an  inadequate representation  for
  galaxies whose luminosity profile slope changes smoothly with radius
  rather than resembling a broken power-law. Thus  we introduce a new,
  alternative     parameter,   and   show    how   this  affects   the
  classification. In fact, we   find  evidence for an   `intermediate'
  class of galaxies, that cannot unambiguously be classified as either
  core  or power-law galaxies, and which  have central cusp slopes and
  absolute magnitudes intermediate between those of core and power-law
  galaxies.  It is unclear at present, however, whether these galaxies
  make  up  a  physically distinct class   or  whether distance and/or
  resolution  effects   cause them to  loose  their  distinct  core or
  power-law characteristics.
\end{abstract}



\ifemulate\else
   \clearpage
\fi


\section{Introduction}
\label{sec:intro}

Ever since the seminal  study by Davies \etal  (1983) it is known that
elliptical galaxies can be divided in two distinct classes. The bright
ellipticals  ($M_B \lta -20.5$) generally have  a boxy appearance, are
pressure  supported,  and often show some   form of radio and/or X-ray
activity.  Low luminosity  ellipticals, on the  other hand, often have
disky isophotes, are  rotationally   supported,  and   generally  lack
significant amounts of activity (Bender 1988; Bender \etal 1989; Nieto
\etal  1988).  Before the Hubble  Space  Telescope (HST) came  online,
very little was  known about the nuclear  properties of these systems.
Based on ground-based observations, with  a typical seeing of $\gta 1$
arcsec,  ellipticals were considered  to  have constant density  cores
with a size that correlates with global galaxy properties (e.g., Lauer
1985; Kormendy 1985).   High-resolution imaging surveys with  the HST,
however, revealed that  the central brightness profiles  of early-type
galaxies are   well  approximated   by  a power-law,     $I(r) \propto
r^{-\gamma}$   with $\gamma > 0$ (e.g., Lauer \etal 1991, 1992; Crane
\etal 1993; Ferrarese \etal 1994; Kormendy \etal 1994;  Lauer  \etal 
1995; Carollo \etal 1997a).  Furthermore, the distribution of $\gamma$
was  found  to  be  bimodal and  to  follow the   dichotomy  of global
properties inferred from ground-based  observations (Jaffe \etal 1994;
Lauer \etal  1995;  Faber \etal  1997) and  to be correlated  with the
global  X-ray  emission (Pellegrini   1999).  Bright  ellipticals have
luminosity profiles  that are well fit  by  a double  power-law with a
break radius of typically a  few hundred parsecs.  The inner power-law
slope  has values of $0  < \gamma \lta  0.3$.  This contrasts strongly
with the low-luminosity  ellipticals whose luminosity profiles  lack a
clear break radius and  have central  power-law profiles with  $\gamma
\sim 0.8$ on average (Jaffe \etal 1994; Ferrarese \etal 1994; Lauer
\etal  1995; Byun \etal  1996; Gebhardt \etal 1996; Faber \etal
1997; Quillen, Bower \& Stritzinger 2000).  Following the nomenclature
of Lauer \etal (1995) we refer to  the former as  core galaxies and to
the latter as power-law galaxies.

In addition to the central density cusps, the images from the HST have
revealed that a    large   fraction of  early-type   galaxies   harbor
significant amounts of dust ($\sim 10^3 -  10^7 \Msun$), either in the
form of a  nuclear disk, or of a  more complex,  filamentary or patchy
morphology (Jaffe  \etal 1994; Lauer \etal  1995; van Dokkum  \& Franx
1995).  In several   cases,     nuclear {\it stellar}     disks   with
scale-lengths as small as $\sim 20$ pc, are found (e.g., van den Bosch
\etal 1994; van den Bosch, Jaffe \& van der  Marel 1998; Scorza \& van
den Bosch 1998).   Both the gaseous  and stellar disks have been shown
to be powerful tools  for determining the  central densities  of their
parent  galaxies, and for detecting  massive  black holes (Harms \etal
1994; Ferrarese,  Ford \& Jaffe 1996; van  den Bosch \& de Zeeuw 1996;
Macchetto \etal 1997; Bower \etal 1998; Cretton \& van den Bosch 1999;
van der Marel \& van den Bosch 1999).  Mainly because of these nuclear
disk structures the    number of galaxies   with solid  detections  of
massive black holes (BHs) has increased strongly  over the past years,
and  we have reached  the point  where we can    start to address  the
demographics of BHs (see Kormendy \&  Richstone 1995, Ford \etal 1998,
Ho 1999, van der Marel 1999a, and Macchetto 1999 for recent reviews).

Black holes  also play  an  important  role   in connection  with  the
observed central cusps.  Two   different scenarios have been  proposed
that  rely on  the  presence of  massive  black holes   to explain the
correlation between cusp slope and luminosity. Faber \etal (1997) have
suggested that luminous galaxies  are  the result of  mergers, whereby
BHs in the progenitors  sink to the  center and  excavate a core  (see
also   Begelman, Blandford  \& Rees   1980; Makino  \& Ebisuzaki 1996;
Quinlan 1996; Quinlan \& Hernquist  1997).  Van der Marel (1999b),  on
the other hand, starts with the  assumption of constant density cores,
and shows that  the adiabatic growth of central  BHs  can create cusps
whose slopes correlate with luminosity as observed.  Thus, whereas the
origin  of the relation between  cusp slope, luminosity and black hole
mass is still unknown, it has become apparent that massive black holes
are omnipresent in galactic nuclei.

Despite the enormous wealth  of  new data, we are   still far  from  a
proper understanding of the  structure  and formation of the   central
regions of early-type galaxies. Not only have the new HST observations
of these systems changed our view of their  central regions, they have
also raised several new problems:

\begin{itemize}

\item  Is the bimodality   in cusp  slopes  real?  Data  on  elliptical
  galaxies with kinematically  distinct cores seems  to suggest a much
  more continuous transition (Carollo \etal 1997a).

\item What is the origin  of the correlation   between cusp slope  and
  global characteristics of early-type galaxies?

\item What fraction of  early-type  galaxies contain nuclear disks  of
  stars or dust/gas,   and how are these  structures  related to other
  properties of their parent galaxies?

\item Is  there a relation between the central structure of  galaxies
  and the presence or absence   of nuclear activity?

\item  How  does the   mass of   a  central BH   correlate  with other
  properties as bulge  mass and cusp slope,  and what role do BHs play
  in determining the properties of the central regions of galaxies?

\end{itemize}

In order to address these  questions one  needs  a large and  unbiased
sample of   early-type galaxies that have    been imaged at comparable
resolution.  On  the order of 80 early-type  galaxies have been imaged
with the HST,  most of them  before the refurbishment mission. We have
obtained HST images with WFPC2 of the central regions of 67 early-type
galaxies as part  of our HST snap  shot program \# 6357.  This roughly
doubles the total number of  early-type galaxies that have been imaged
with the HST.   Together with the data taken  so far  and with similar
samples of spiral  bulges (Phillips \etal  1996;  Carollo \etal 1997b;
Carollo,  Stiavelli \& Mack 1998)  and  radio galaxies (de Koff  \etal
1996; McCarthy \etal 1997;  Martel  \etal 1999; Verdoes Kleijn   \etal
1999) these   images  provide a   data base for  the  investigation of
correlations  of properties   such   as  dust,  metallicity,   colors,
activity, and nuclear structure.

In this paper, the first in a series, we present the data and describe
the   luminosity  profiles  and  isophotal structures.   More detailed
analyses  of  the  dust, stellar  disks,    and the central  parameter
relations are deferred to future papers.

\section{The Sample} 
\label{sec:sample}

The sample is  compiled from  the Lyon/Meudon Extragalactic   Database
(LEDA)\footnote{www-obs.univ-lyon1.fr}, by  selecting  all  early-type
galaxies with  radial velocities  less  than $3400 \kms$,  an absolute
$V$-band  magnitude less than  $-18.5$, and absolute galactic latitude
exceeding 20 degrees (to minimize the effects of galactic extinction).
In total 130 galaxies made  it into the  sample, of which 68 have been
observed successfully by HST. Thus, although our sample is by no means
complete, it  has  been homogeneously selected.  One galaxy, UGC~5467,
turned  out to be    a spiral galaxy  and will   be discarded in  what
follows.   Note  that, although we  have attempted  to discard obvious
duplicates (WFPC2) from  our sample, several  of our objects have been
observed in   other  HST programs  since   the  time  our  sample  was
selected. In most cases, however, a different filter was used than for
the data presented here.

Global properties  of the 67 galaxies   in our sample,  taken from the
LEDA (see Paturel \etal  1997), are listed in  Table~\ref{tab:sample}.
Absolute  magnitudes (column~2) are based  on  the distances listed in
column~4, and based on the Virgo-centric infall corrected heliocentric
velocities  assuming $H_0 =  80  \kmsmpc$ (see  Paturel \etal 1997 for
details).  In  addition, we list the  1.4 GHz radio fluxes, taken from
the NRAO  VLA Sky  Survey  (NVSS; Cordon \etal   1998),  as well  as a
far-infrared magnitudes defined as
\begin{equation}
\label{mfir}
m_{\rm FIR} = -2.5 \, {\rm log}(2.58 f_{60} + f_{100}) + 14.75
\end{equation}
(cf. de Vaucouleurs  \etal 1991). Here $f_{60}$  and $f_{100}$ are the
q60 $\mu$m and 100 $\mu$m IRAS fluxes measured by  one of us, adopting
a  similar approach to that taken by Knapp \etal (1989).

\section{Observations and Data Reduction}
\label{sec:obs}

\subsection{Observing strategy}
\label{sec:strategy}

We used the WFPC2  to obtain broad band images  of the galaxies in our
sample, using the  F702W filter.  This filter,  centered on $\lambda =
6997$\AA$\;$ and   with a  FWHM of   $\sim 1481$\AA  , was   chosen to
compromise between maximizing the number of photons and minimizing the
effects  of dust  on the  observed  morphologies.   The nuclei  of the
galaxies were  centered  on the  Planetary    Camera CCD (PC),   which
consists  of  $800 \times 800$  pixels   of $0.046\arcsec \times 0.046
\arcsec$ each. All exposures were taken  with the telescope guiding in
fine   lock,    yielding a  RMS   telescope    jitter  of   $\sim   3$
milli-arcseconds. More detailed information on  the WFPC2 can be found
in Biretta \etal (1996).

Previous  HST/WFPC2 observations  of ellipticals  have shown that  the
power-law galaxies have typically a central  surface brightness in the
$V$-band of $\mu_0 \approx 14.5$, whereas core-galaxies typically have
$\mu_0 \approx 16.5$. Our aim is to  obtain a signal-to-noise (S/N) of
20 or larger in the center of the galaxies.  This prompted us to use a
total exposure time of 1000 seconds  with the analogue-to-digital gain
set to 13.99 electrons/DN  (DN is the  number  of counts).   With this
setting the read-out noise  is $7.02$ electrons.  To facilitate cosmic
ray  removal, and to   guard against saturation in  the  presence of a
bright nuclear  point source, each exposure was  split in two separate
exposures of 300 and 700 seconds respectively.

\subsection{Data reduction}
\label{sec:reduction}

The images are calibrated by the HST calibration `pipeline' maintained
by  the Space Telescope   Science  Institute.  The standard  reduction
steps    include  bias subtraction,    dark   current  subtraction and
flat-fielding, and are described in detail by Holtzman \etal (1995a).

Subsequent   reduction is done using  standard  IRAF tasks.  Using the
data quality files  and the WFIXUP  task, bad pixels are corrected  by
means of a linear one-dimensional interpolation. The alignment between
the  300  and 700  exposures is determined   by comparing the isophote
centers from initial  analyses   of the  separate exposures.  If   the
misalignment  is   significant  the  300    sec  exposure  is  shifted
appropriately, using  linear interpolation.  A difference-image of the
700 sec and (shifted) 300 sec images is used to  check the accuracy of
the alignment.  Next,  the two images  are  combined with simultaneous
removal of cosmic rays, using the IRAF task IMCOMBINE.  The cosmic ray
removal  and  bad pixel  correction  is checked  by  inspection of the
residual  images between the cleaned, combined   image and each of the
two  original frames.  Sky levels are  determined from  empty areas in
the WF images, and accordingly subtracted.   The sky-level is found to
be  small, typically $\sim  1$ DN.  The  count rates in the broad-band
F702W images are calibrated to magnitudes in the Landoldt $R$-band, as
described  in Holtzman \etal (1995b),  and assuming a $V$-$R$ color of
$0.68$.  The magnitudes thus   derived are checked using   the inverse
sensitivity and the zero point of the  magnitude scale provided by the
HST pipeline, whereby it is assumed that ellipticals have the spectrum
of a K-III giant. The consistency is found to be better than 0.01 dex.

The 300 and  700 sec exposures  are reduced separately as well. Cosmic
rays are removed using the IRAF task COSMICRAY.  These images are used
to check the consistency of structures, for obtaining estimates of the
uncertainties due to random noise, and for  cases where saturation has
affected the 700 sec exposure (see \S~\ref{sec:lumprof}).

Despite  the high  spatial  resolution  of the  refurbished HST,   the
central few tenths of an arcsec of the galaxies  are influenced by the
effects    of point-spread function   (PSF)  smearing.  Because of the
stability of the HST PSF, these effects can  be corrected for by means
of deconvolution with the Richardson-Lucy  (RL) algorithm (Lucy 1974).
Based on tests with model galaxies  (see \S~\ref{sec:models} below) we
deconvolve the reduced  300 sec, 700  sec and combined  images of each
galaxy with 40 iterations of the RL algorithm. For three galaxies with
faint and shallow surface brightness  profiles (NGC~3613, NGC~4168 and
NGC~4365), the RL-algorithm with  40 iterations ends up amplifying the
noise too strongly  for an accurate analysis   of the central  surface
brightness profile,  and only 15 iterations are   used in these cases.
For each galaxy  we use a  separate model PSF  $(5.7'' \times 5.7'')$,
created with the TINYTIM  software package  (Krist 1992), centered  on
the location  of the galaxy's center in  the PC CCD\footnote{The exact
shape of the  HST PSF depends on  its location  in the focal  plane.}.
Contour plots of the deconvolved images are presented in the Appendix.

\section{Data Analysis}
\label{sec:analysis}

The main goal of this paper is to analyze  the isophotal structure and
luminosity profiles of the galaxies. For the isophotal analysis we use
the  isophote fitting task  ELLIPSE in  IRAF.   For each isophote, the
center, ellipticity  ($\epsilon$),  and position angle  ($\theta$) are
computed.  In addition,  the third and  fourth order deviations of the
isophote  from a pure ellipse  are determined.  These are described by
the amplitudes ($a_n$ and  $b_n$  with $n=3,4$)  of  the sin and   cos
$3\theta$ and $4\theta$ terms of  the following Fourier expansion (see
Jedrzejewski 1987):
\begin{equation}
  \frac{\delta r(\theta)}{r(\theta)}= \sum_{n=3}^{4} \left[ a_n
    \sin{(n\theta)} + b_n \cos{(n\theta)} \right]
\end{equation}
For  a perfectly elliptical  isophote these coefficients are all equal
to  zero. Typically the absolute values  of $a_n$ and $b_n$ are small,
rarely  exceeding  $0.02$.   The $b_4$   coefficient   is of   special
interest, since  positive (negative) $b_4$  values correspond to disky
(boxy) isophotes.

In order to parameterize the luminosity  profiles they are fitted with
a  double power law, known  as the ``Nuker''-law  profile (Lauer \etal
1995):
\begin{equation}
\label{nukerlaw}
I(r)=I_b \; 2^{(\beta-\gamma)/\alpha} \left (\frac{r}{r_b} \right
)^{-\gamma} \left [1 + \left (\frac{r}{r_b} \right )^{\alpha} 
\right]^{(\gamma-\beta)/\alpha}
\end{equation}
which has $I(r)  \propto r^{-\beta}$ at $r^{\alpha} \gg r_b^{\alpha}$,
and $I(r) \propto r^{-\gamma}$ at  $r^{\alpha} \ll r_b^{\alpha}$.  The
parameter $\alpha$ describes the sharpness of the break from the outer
to the inner power law.   The  break-radius, $r_b$,  is the radius  of
maximum curvature  in log-log coordinates,   and $I_b$ is  the surface
brightness at $r=r_b$.  The Nuker law has been shown to accurately fit
the central   luminosity profiles of  early-type galaxies  (e.g., Byun
\etal 1996).  Note that  equation~(\ref{nukerlaw}) is only intended to
fit the  luminosity profiles in  the  inner $\sim  20''$ of early-type
galaxies.   For typical  fitted values   of  $\beta$ there  must be  a
further  downturn   in the  profile  at  larger  radii  for  the total
luminosity to be finite.

\subsection{Model galaxies}
\label{sec:models}

In  order  to test the   accuracy  of our  reduction   and analysis we
constructed a large  set of model  galaxies.  Each model  galaxy has a
luminosity profile   of the  form of    equation~(\ref{nukerlaw}), has
constant ellipticity and position  angle, and has perfectly elliptical
isophotes (i.e., $a_n=b_n=0$ for $n \geq 3$).  A fake image is created
by integrating the model   surface brightness distribution   over each
pixel, and by adding photon and read-out noise.  Pixel sizes are taken
the same as for the  true data.  The frame  size  of $200 \times  200$
pixels  is smaller  than  the  actual  HST  images, but satisfies  the
requirements for our testing purposes.

The resulting image, refered  to as `unconvolved' image,  is convolved
with the HST PSF (the `convolved' image), and subsequently deconvolved
using $N_{\rm iter}$ iterations of the RL algorithm (the `deconvolved'
image).    The PSFs used  for  the  convolution  and deconvolution are
constructed with TINYTIM, and differ slightly from each other in order
to  mimick    errors in the    exact  PSF shape. The   unconvolved and
deconvolved model images are analyzed in the same  way as the data: an
isophotal analysis  is performed, and we fit  the Nuker-law profile to
the major axis luminosity  profile    determined from the     isophote
fitting.  The  deconvolution  is tested with   10, 15,  25,  40 and 80
iterations. The best results were obtained for $N_{\rm iter} = 40$, in
good  agreement with  Lauer  \etal (1998).  However, for galaxies with
shallow central  surface brightness profiles $N_{\rm iter}=15$ already
suffices  to accurately recover    the  unconvolved image,  and   more
iterations only amplify the noise.

In total we have  constructed $15 \times  50$ model galaxies. Each  of
the 15 sets of model galaxies is characterized by a value of $\gamma$,
ranging  from $\gamma=0$ to $\gamma=1.4$ in  steps of $0.1$.  For each
set 50 model galaxies are created  whose luminosity profile parameters
are varied at the few  percent level.  The ellipticity and orientation
of each model galaxy are chosen  randomly from the intervals $[0,0.5]$
and $[-90,90]$, respectively. In addition,  the center of the  galaxy,
in fractional pixels, is drawn randomly as well.

In  Figure~\ref{fig:modeliso} we plot  the isophotal parameters of one
of the model  galaxies   with $\gamma=0.7$.  Even  though   this model
intrinsically has $\epsilon = 0.26$ over its  entire radial range, the
ellipticities found  by the isophote-fitting routine are significantly
lower  for $r   \lta 0.2''$,  even for   the unconvolved image  (solid
triangles).  In addition, the higher-order parameters become noisy and
reveal fluctuations  inside this radius.  These  effects owe mainly to
the discrete pixel sampling   and the sub-pixel interpolation used  by
the isophote-fitting routine, and indicate that one  can not trust the
isophotal parameters  for  $r \lta  0.2''$ ($\sim 4$  pixels).   Note,
however, that the isophotal parameters of the deconvolved model galaxy
are remarkably  similar to those of  the unconvolved model, indicating
that the deconvolution   routine   used does not  introduce   spurious
artefacts.

In order to examine  the accuracy with  which we can recover the Nuker
parameters of the luminosity  profile used as input  for the model, we
first focus on the unconvolved images.  For each model $i$ we compute
\begin{equation}
\label{deltamu}
\Delta \mu_{i}(r) = \mu_{0,i}(r) - \mu_{{\rm iso},i}(r)
\end{equation}
with  $\mu_{0,i}(r)$ the input  surface  brightness profile (in  magn.
arcsec$^{-2}$)   of equation~(\ref{nukerlaw}),     and      $\mu_{{\rm
iso},i}(r)$  the major axis  lumninosity  profile determined from  the
isophotal analysis of the unconvolved image.  For  each of the 15 sets
of model galaxies   (that have the  same  value of  $\gamma$), we then
compute the average    $\langle   \Delta  \mu \rangle(r)$  and     the
corresponding    standard deviation $\sigma_{\Delta\mu}(r)$, where the
average is over the 50 models in each set.

Figure~\ref{fig:deltamu}   plots $\langle  \Delta  \mu \rangle(r)$ and
$\sigma_{\Delta\mu}(r)$  for 8 of the 15  sets. For $r \gta 0.2''$ the
luminosity  profiles   derived  from  the isophotal  analysis   are in
excellent agreement with the   intrinsic profile.  For  smaller radii,
however, the error increases   strongly, especially for the   profiles
with  steep central cusp  slopes.    Even more worrysome  is  that the
average   residual $\langle \Delta   \mu \rangle(r)$ strongly deviates
from zero  at small radii, implying  that the main contribution to the
error is     of systematic rather   than  random  nature.    This is a
reflection of the problems with    the isophotal analysis due to   the
discrete,   pixelized nature    of the     data   and the    sub-pixel
interpolation.   As   Jedrzejewski (1987) already    pointed out,  the
sub-pixel interpolation  scheme used  can introduce spurious artefacts
especially when the  gradients  in the  luminosity  profiles are large
(i.e., when they are steeply cusped).

Clearly, when  determining  the  intrinsic   central cusp  slopes   of
galaxies one can not  use the  surface brightness profiles  determined
from the isophotal  analysis.  We therefore use a  different approach.
For    $r   \geq   0.2''$   we     fit   the  Nuker  law      profiles
(equation~[\ref{nukerlaw}])  directly to  the   major axis  luminosity
profile determined  from the isophotal   fit to the  image.  For  $r <
0.2''$, however, we extract the actual  pixel values of the image from
the data, which we compare to the model predictions given by
\begin{equation}
\label{pixelvalues}
I_{\rm fit}(x,y) = \int \!\!\!\!\!\! \int\limits_{\rm pixel} I(\hat{m})
\; {\rm d}\hat{x} \; {\rm d}\hat{y},
\end{equation}
with $I(r)$ the Nuker law profile and
\begin{equation}
\label{msquared}
\hat{m}^2 = \hat{x}^2 + \left( \frac{\hat{y}}{1-\epsilon_0} \right)^2.
\end{equation}
Here  $(\hat{x},\hat{y})$ are the  coordinates centered  on the galaxy
center $(x_0,y_0)$ with the $x$-axis  rotated  by an angle  $\theta_0$
with respect  to the  pixel  coordinate system, and   $\epsilon_0$ and
$\theta_0$   are the ellipticity and  position  angle of the isophotes
inside $0.2''$, respectively.  We  make   the assumption that   $x_0$,
$y_0$,  $\epsilon_0$, and $\theta_0$  are  all constant inside $0.2''$
and we  compute  their  values  as  the average  of  the corresponding
isophotal  parameters in the radial  interval from $0.2''$ to $0.4''$.
In  cases where  the    isophotes   in this   radial  range   are  not
representative of   the galaxy, e.g., distorted   by dust, a different
more appropriate radial range is chosen.

The advantage   of this method  is  that by using  the  observed pixel
values to determine the  best fit Nuker profile we  do not suffer from
any  problems  with sub-pixel  interpolation  which  hampers  a proper
isophotal analysis at small radii.  The disadvantage, however, is that
if the actual isophotal parameters  vary  with radius inside  $0.2''$,
this  will introduce systematic  errors in the  parameters of the best
fit Nuker law profile. However,  detailed tests, described below, show
that these errors are much smaller  than those stemming from sub-pixel
interpolation. Note  that a somewhat  similar scheme was  used by Byun
\etal (1996), who used the total integrated light inside $0.1''$ as an
additional constraint for the Nuker-law fitting procedure.

Figure~\ref{fig:modelsb} illustrates  the advantage of our new method.
The solid lines  correspond   to the intrinsic  major  axis luminosity
profile of the same model  as in Figure~\ref{fig:modeliso} (i.e., with
$\gamma=0.7$).  The upper profiles with  the filled circles correspond
to the luminosity profiles determined from the isophotal analysis.  As
is evident,  inside $\sim 0.1''$  the sub-pixel  interpolation used by
the isophotal analysis introduces  systematic errors.  Note again that
the results  for the unconvolved  (left panels) and deconvolved (right
panels) model  images  are  virtually identical,  indicating that  the
deconvolution procedure used   can  accurately  correct for   the  PSF
smearing.  The  lower profiles with  open symbols, offset by 1.5 magn,
correspond to the luminosity  profile determined with the  new method.
The open circles ($r \geq 0.2''$) are the luminosities determined from
the isophotal analysis.  The  open  squares correspond to the   actual
pixel values  of  the  model image  inside  $0.2''$\footnote{Since  in
general the (center of the) pixels do  not lie exactly along the major
axis, we make use of the following way to  visualize the fit residuals
in the luminosity profile plot: We plot the pixel values at the radius
$r_{plot}$  defined such   that $I(r_{plot})=I_{fit}(x,y)$, i.e.,  the
difference between the  pixel   value and  the Nuker-law profile    at
$r_{plot}$ is equal to the  difference between the actual pixel  value
and the prediction $I_{fit}(x,y)$ of the  best-fit model. This way the
residual  is accurately  visualized.}.  As is   evident from the lower
panels, this new method allows us  to recover the intrinsic luminosity
profile with high accuracy down to very small radii.

In  order to determine the  parameters  of the  Nuker law profile that
best  fits the data we  use  a $\chi^2$-minimization technique.  For a
given set of parameters ($\alpha$,  $\beta$, $\gamma$, $r_b$,  $I_b$),
we  determine $\chi^2 \equiv \chi_1^2    + f^{-2} \, \chi^2_2$.   Here
$\chi^2_1$ and $\chi^2_2$ are  the $\chi^2$-values of  the pixel-value
fitting  inside $0.2''$ and the  luminosity profile fitting at $r \geq
0.2''$,  respectively.   As errors we use  Poisson  statistics for the
pixel values and  the nominal errors  given by the isophotal analysis,
respectively. In practice, these latter  errors are underestimated, as
is evident  from the fact  that the normalized  $\chi^2_2$ of the best
fit model is typically   significantly more than unity. We   therefore
introduce the parameter $f$, which  scales the nominal error given  by
the isophotal analysis, and which thus sets the relative contributions
of $\chi^2_1$ and $\chi^2_2$ to the total $\chi^2$ of the fit.  In the
left  panel of Figure~\ref{fig:fchi}   we plot the average  difference
between  the intrinsic cusp steepness    $\gamma_{\rm model}$ and  the
best-fit Nuker-law profile $\gamma_{\rm fit}$, as  function of $f$ for
the set of models  with $\gamma=0.5$. The averages  are taken over all
50 model galaxies  in this set.  As expected, the  fits are better for
$f>1$.  A  value of $f \simeq 5$  yields values for $\gamma_{\rm fit}$
in good agreement with $\gamma_{\rm model}$.  For $f \lta 1$, too much
relative weight  is given  to the  pixel  values  inside $0.2''$,  and
$\gamma_{\rm fit}$  is   too large. For   too large  $f$,  $\chi^2$ is
completely dominated by the isophotal luminosities outside $0.2''$ and
the central cusp slope becomes poorly constrained,  as is evident from
the large errorbars. In what follows we therefore adhere to $f=5$, for
which in general we find a  normalized $\chi^2$ of  the best fit model
close to unity.

We have constructed luminosity  profiles for all 750 deconvolved model
images using  our new method, and   determined the best  fit Nuker law
parameters using the  $\chi^2$-minimization technique  described above
with $f=5$. In Figure~\ref{fig:fchi} we plot $\gamma_{\rm fit}$ versus
$\gamma_{\rm model}$  for all 750 models.   As is evident, this method
of analyzing luminosity  profiles allows an accurate  determination of
the central  cusp slope, and  we use  the  same method  to analyze the
luminosity profiles of the actual HST images.

\section{Results}
\label{sec:results}

\subsection{Isophotal Analysis}
\label{sec:isoph}

For each galaxy there are six reduced  images: the non-deconvolved and
deconvolved  300 sec,  700 sec, and    combined images.  An  isophotal
analysis on each  of these six   images is performed.   All images are
checked for contamination  such  as cosmic  rays  that have  not  been
properly  removed,  foreground    stars,  dust filaments,  and  bright
globular clusters.  If necessary, these   objects are masked out.  Ten
galaxies are so severely  influenced by dust that meaningful estimates
of  their luminosity  profiles and  isophotal  structures  can not  be
obtained (all  galaxies with filamentary dust of  ``type'' III as well
as  NGC~4233  and  NGC~4494,  which  have a   large   dust disk,   see
\S~\ref{sec:dust}  below).  In most  of  the  following  discussion we
discard  these  galaxies and  focus on the  sample  of 57 galaxies for
which the HST images provide a  meaningful picture of the structure of
the stellar component. In what follows we  refer to this sub-sample as
the {\it unperturbed} sample.

Isophotal parameters  are   determined down  to   $r  =  0.03''$,  and
presented in  the Appendix. Note that,  although we plot the isophotal
parameters down  to $r  = 0.03''$, only  the   parameters for $r  \gta
0.2''$  are  considered  for further  analysis.   For comparison,  the
isophotal  parameters of  the   deconvolved image (open   circles) are
overplotted  with   the   ones of   the  non-deconvolved image  (solid
triangles).   At larger radii,  unaffected by convolution effects, the
latter data, which  is less noisy,  nicely complements the deconvolved
data.

Previous high-resolution imaging with   the HST has revealed that  the
central regions of early-type  galaxies are very complex environments. 
This is strikingly confirmed with our sample, and is clearly reflected
in the isophotal  parameters: the isophotal  shapes can  vary strongly
with radius and are very different from  one galaxy to the other. This
makes it  very hard  to  categorize the central regions  of early-type
galaxies in a small  number of  classes.  It  has become  customary to
classify ellipticals as either disky or boxy, depending on the sign of
the $b_4$-parameter.  This generally works well for the outer parts of
ellipticals, where the changes in the isophotal shape are only modest.
However, the  central regions reveal a vastly  more complex behavior. 
A single galaxy can have isophotes that  change from disky to boxy and
back to disky  all within the  inner  $10''$, even  when there  are no
signs of dust.  Clearly, a bimodal classification does not suffice to
describe the bewildering variety of morphologies apparent in the inner
$\sim 10''$.   Another approach  is   therefore employed: Rather  than
classifying galaxies as  either disky or  boxy a simple letter coding,
referred to as    {\it isophotal code},    is  used to  describe   the
individual   isophotal shapes of  each galaxy.    The letters indicate
regions with  common  isophotal  shape,  starting from the   center at
$r=0.2''$ out to  $r \simeq 25''$, where   {\it d}, {\it b},  {\it 0},
{\it x} and {\it   ?} refer to disky,  boxy,  neither disky  nor boxy,
undetermined  due to  dust,  and undetermined   due to small   surface
brightness gradient, respectively.  As  an example, the isophotal code
{\it x0d} for NGC~2592 indicates  that  for small radii the  isophotal
shape could not be determined  due to dust (in this  case a small dust
disk, see~\S~\ref{sec:dust}), followed by a region where the isophotes
are elliptical, and which become disky  at larger radii.  As a general
rule, a galaxy region is classified as disky (boxy),  if $b_4 > +0.01$
($b_4 < -0.01$) for  several   consecutive isophotes while  the  other
high-order   parameters are not   significantly  different from  zero. 
Exceptions  are  misaligned structures which  also  result in non-zero
$a_4$   values.   A  more detailed    description of these  intriguing
structures is given in \S~\ref{sec:misdisk}.

The isophotal codes for each    galaxy are  listed in column~(5)    in
Table~\ref{tab:isoparameter}.    Clearly,   this encoding  is somewhat
subjective.  Nevertheless, it is  well suited to describe the  complex
and diverse isophotal  behavior of the  galaxies.  In  addition to the
isophotal  code, the  median  ellipticity $\langle  \epsilon  \rangle$
calculated over the radial interval from $1.0''$ to $10.0''$ is listed
in Column~(8) of Table~\ref{tab:isoparameter}.  Note that these median
values may not  be very  meaningful   for galaxies that  have  rapidly
changing  ellipticities. Large variations may  be  real or may reflect
the presence of  dust, and are easily  identified  by their relatively
large errors ($\Delta \langle \epsilon \rangle \gta 0.1$).

\subsection{Luminosity Profiles}
\label{sec:lumprof}

Luminosity profiles  along   both  the  major and minor    axes of the
deconvolved  images  are determined and   subsequently fitted with the
Nuker-law  profile as described in  detail in \S~\ref{sec:models}.  In
some cases a deviation  from  equation~(\ref{nukerlaw}) is visible  in
the inner $\sim 0.1''$ as excess light,  and pixels inside $0.1''$ are
ignored in the  fitting procedure. These (largely unresolved) `nuclei'
are  discussed in more detail   in \S~\ref{sec:nuclei}.  In the  cases
where a  central dust disk is present,  we exclude any data inside the
major  axis  radius of  the dust  disk   from the  fit.  The best  fit
parameters      are      listed    in      Tables~\ref{tab:NukerMajor}
and~\ref{tab:NukerMinor} for  the major  and minor axes, respectively.
Plots of  the luminosity profiles are  presented in the Appendix, with
the best fit  Nuker law profiles  overplotted (solid lines).  In cases
where the  luminosity profiles do not reveal  a clear change  in slope
over the entire radial range (NGC~4494 and NGC~3377 along the minor
axis) the observed luminosity profile is fitted
by a single power law:
\begin{equation}
\label{powerlaw}
I(r) = I_0 r^{-\beta}
\end{equation}
For these galaxies, the values listed for $I_b$ correspond to $I_0$ in
Tables~\ref{tab:NukerMajor} and~\ref{tab:NukerMinor}, and no values
for $\alpha$, $\gamma$ and $r_b$ are given.

\subsubsection{Cusps versus Cores}
\label{sec:class}

As discussed in \S~\ref{sec:intro}, HST images  have revealed that all
early-type galaxies have luminosity profiles with central cusps.  More
importantly, it  was  found that the  distribution  of cusp slopes  is
bimodal  and in  support of the  dichotomy  of early-type galaxies  as
inferred from their isophotal   and kinematical structure.   In  their
study of the central parameter relations of early-type galaxies, Faber
\etal  (1997;  hereafter F97) classified  all  galaxies with $\gamma <
0.3$ and $r_b \geq 0.16''$ as core  galaxies.  All other galaxies were
classified as power-law galaxies. It is important to realize that such
classification scheme suffers from distance effects. Galaxies at large
distances will  have  break radii  $r_b$  that  are smaller  than  the
resolution limit of  the  simulations. Consequently, the  central cusp
slopes of the observed luminosity profiles of these  systems are to be
compared  to  $\beta$,  and not  $\gamma$,   of less distant  systems.
Similarly, galaxies that are  much closer than  the main population of
systems   might reveal central  luminosity  profiles that deviate from
the  simple double-power law.  Indeed, F97  pointed  out that both M31
and M32 are  poorly  fit by   the Nuker-law  profile, and   that their
classification would have been different had they been at the distance
of Virgo.

With these caveats in mind, we now follow F97 and attempt to classify
the galaxies in our unperturbed sample as either `core' or
`power-law'. However, when resorting to the same classification scheme
as used by F97 we are confronted with another shortcoming concerning
luminosity profiles having small $\alpha$. Such profiles occur in our
sample for two reasons: because of the distance-problem alluded to
above, or because they intrinsically have luminosity profiles for
which the slope changes smoothly with radius. Some galaxies in the
latter category can still be fit by a Nuker-law profile, while for
others, even small $\alpha$ Nuker-law profiles fits cannot accomodate
all profile features.  In all of these cases we generally find
best-fitting Nuker-law profiles with $\gamma \simeq 0$.  However,
$\gamma$ is only a proper representation of the gradient in the
luminosity profile at radii $r$ for which $(r/r_b)^{\alpha} \ll 1$.
For small $\alpha$, however, $(r/r_b)^{\alpha}$ can be close to unity,
even when $r$ is significantly smaller than $r_b$, and $\gamma$
therefore becomes a measure of the gradient of the luminosity profile
at radii much smaller than the actual resolution limit.  Since one has
to be sceptical about basing a galaxy's classification on extrapolated
quantities, we complement our classification scheme by introducing the
parameter
\begin{equation}
\label{gamma_prime}
\gamma' \equiv - \left[{{\rm d}\log{I} \over {\rm d}
    \log{r}}\right]_{r=0.1''} = {\gamma + \beta \left( {r' \over r_b}
  \right)^{\alpha} \over 1 + \left( {r' \over r_b}
  \right)^{\alpha}}
\end{equation}
with $r' =  0.1''$.  Thus, $\gamma'$  is determined directly from  the
best-fit Nuker-law, and is a measure of the gradient of the luminosity
profile  at $r=0.1''$.   The values  of $\gamma'$ along  the major and
minor    axes of      the     luminosity profiles  are     listed   in
Tables~\ref{tab:NukerMajor} and~\ref{tab:NukerMinor}, respectively.

In  the upper  panels of  Figure~\ref{fig:gamma} we   compare the cusp
slopes of the luminosity profiles along the major and minor axes. Both
$\gamma$   and $\gamma'$ give very   similar  results along both axes,
indicating  that our method of  using the actual observed pixel values
inside $0.2''$ to constrain the Nuker-law fit  yields a robust measure
of the  central cusp slope.   Comparing $\gamma$ to $\gamma'$ for both
the major and  minor   axis  luminosity  profiles (lower  panels    of
Figure~\ref{fig:gamma})  yields a more  complicated pixture.  Galaxies
with $\gamma > 0.5$ in  general have $\gamma' \simeq \gamma$. However,
for $\gamma  < 0.3$ about half of  the galaxies  have $\gamma' < 0.3$,
and the other half has $\gamma' > 0.3$. We therefore use the following
classification scheme,   based   on  both  $\gamma_{\rm     maj}$  and
$\gamma'_{\rm maj}$\footnote{our   results would  be identical had  we
used $\gamma_{\rm min}$  and $\gamma'_{\rm min}$}: core  galaxies have
$\gamma'_{\rm maj} < 0.3$, power-law galaxies have $\gamma_{\rm maj}
\geq  0.3$, and   galaxies   with $\gamma'_{\rm   maj} \geq   0.3$ and
$\gamma_{\rm    maj}  <  0.3$   are    provisionally  refered  to   as
`intermediate' galaxies. Note that  this classification scheme is more
conservative in designating galaxies  as  core galaxies as the  scheme
used  by  F97:  In  our  sample, only  $56$  percent  of  the galaxies
classified as core galaxies according to the F97 scheme are classified
similarly with our definition.   The other $44$ percent are classified
as  intermediate. Column~(2) in Table~\ref{tab:isoparameter} indicates
the classification of our galaxies  as either core `$\cap$', power-law
`$\setminus$', or intermediate `)'.    A question mark  indicates that
the galaxy  is not in the unperturbed  sample, and that no attempt has
been made to classify this galaxy.

Figure~\ref{fig:gammahisto} presents histograms of the distributions
of $\gamma$ and $\gamma'$. Although the distribution of cusp slopes
reveals a clear hint for bimodality, consistent with the results of
F97, we find less of a deficit of galaxies with $0.3 \leq \gamma \lta
0.5$.  Most of the galaxies with $\gamma$ in this range, and which are
thus classified as power-law galaxies (e.g., NGC~5576, NGC~5796, and
NGC~5831), have small values of $\alpha$ and no disky isophotes,
similar to the intermediate galaxies.  The intermediate galaxies
themselves follow the same distribution of $\gamma$ as the {\it core
galaxies}, but at the same time occupy the low-$\gamma'$ end of the
{\it power-law galaxies}.  Thus, whereas we confirm the bimodality in
cusp slopes found by F97, we find that $\sim 15$ percent of the
galaxies in our unperturbed sample have luminosity profiles whose
profile slopes change very gradually with radius (reflected by the
small value of $\alpha$), making a classification based on central
cusp slope somewhat ambiguous (see discussion in
\S~\ref{sec:cusp-slopes} below).

\subsection{Dust Properties}
\label{sec:dust}

As discussed in the introduction,  the high resolution images obtained
with  the HST have  strongly increased the  detection  rate of dust in
early-type   galaxies.     Two   different  dust   morphologies    are
distinguished:  disks (or tori) and   filaments (indicated by `d'  and
`f', respectively, in  Table~\ref{tab:isoparameter}).   The amount  of
filamentary dust present is indicated by a Roman numeral, ranging from
I (small traces of dust that hardly influence the isophotal shapes) to
III  (large amounts of dust that  prevent a meaningful analysis of the
isophotes and  luminosity profiles).  Types   II  and III  are  easily
discerned from an inspection of the images by eye, whereas filamentary
dust of class I is only detected  after a more detailed investigation. 
A  detailed  discussion of    the  dust properties  in our   sample is
presented   in Tran   \etal  (2000).  Here  we  merely list  the  main
statistics.

In   10 galaxies (15 percent)  we  find a nuclear    dust disk, with a
diameter ranging from $0.3''$ (NGC~4648) to $12.3''$ (NGC~4233).  Most
remarkably, 8 of these  galaxies are classified as power-law galaxies,
whereas  the two remaining,  NGC~4233  and NGC~4494,  have a dust disk
that is too  large  to be able to   classify the galaxy  based  on its
luminosity profile.   Filamentary  dust   is  found in 19  cases   (28
percent).  In total, we  thus find evidence for dust  in 43 percent of
our  galaxies.  Van Dokkum   \& Franx (1995), analyzing 64  early-type
galaxies imaged with the  HST before the refurbishment mission,  found
dust in  48 percent of  the cases and  pointed out that  a significant
fraction of the dust  will be missed  because of  inclination effects.
They estimated that  $\sim 75$  percent   of the early  type  galaxies
harbor significant amounts of dust.  Our results are in good agreement
with their estimates.

\subsection{The dichotomy amongst early-type galaxies}
\label{sec:dichotomy}

\subsubsection{Isophotal shapes}
\label{sec:isoshapes}

As  mentioned  in the introduction,  early-type  galaxies  come in two
different  classes,  core and    power-law galaxies.  The   latter are
relatively faint ($M_B  \gta -20.5$), typically have disky  isophotes,
and are thought to be rotationally  supported spheroidal systems. Core
galaxies, on the other hand, are relatively bright systems, often with
boxy isophotes,  that are   thought   to be triaxial and    henceforth
pressure-supported.  We now investigate to what extent the galaxies in
our sample obey this  dichotomy.  As discussed  in \S~\ref{sec:isoph},
the isophotal structures of the  central regions are very complex, and
we   introduced  an  isophotal code   (hereafter   IC) to describe the
observed morphologies. The galaxies  can roughly be divided  into four
classes: disky (`d', ICs with $d$ but no $b$), boxy (`b', ICs with $b$
but no $d$), disky-boxy (`db', ICs with both $d$ and $b$), and regular
(`0', ICs with no $d$ or $b$).

In Table~\ref{tab:lumisocorr}  we  correlate this classification based
on the isophotal structure with that based on the luminosity profiles.
We only  consider the  57  galaxies  in our  unperturbed  sample.   In
general, our sample  nicely  confirms the  dichotomy that has  emerged
from previous  studies (e.g., Jaffe  \etal 1994; Lauer \etal 1995): 20
of the 21  disky (`d') galaxies  are classified as power-laws, while 7
of the 9 core galaxies are classified as  either boxy (`b') or regular
(`0').  However, not  {\it all} galaxies  fall in either of these  two
classes.   As  we  indicated  in   \S~\ref{sec:class}, roughly  twelve
percent of  our galaxies classify   as `intermediate' based  on  their
luminosity  profiles.  Remarkably, none  of  these galaxies reveal any
sign of diskiness.  In fact, two of them classify as boxy, whereas the
other four are regular.  As such, these intermediate galaxies are more
similar to core galaxies than to power-law  galaxies.  Of the in total
8 galaxies with both disky  and boxy isophotes  (`db'), 7 classify  as
power-law  galaxies.  In total,  we thus find  that  93 percent of all
galaxies with  disky isophotes are power-law  galaxies.   If one takes
into account that nuclear  disks are much   easier detectable in  core
galaxies than in power-law galaxies, it is clear that  there is a very
strong  correlation between the  diskiness  and central  cusp slope of
early-type galaxies.

Four of the in  total 57 galaxies  in  our unperturbed sample  deviate
from   these  general  trends: NGC~3078   and   NGC~5576 are the  only
power-law galaxies (out of 41) with boxy  isophotes.  NGC~3078 is also
the galaxy with the highest 1.4 GHz flux in our  sample, and reveals a
central    dust disk with  a  diameter  of $\sim   1.4  ''$.  The dust
absorption   combined with the   presence  of a   strong AGN makes the
classification rather uncertain, and it is thus not inconceivable that
NGC~3078 is in fact a core galaxy. NGC~5576, on the other hand, has no
detectable dust or strong radio  emission. However, with  $\gamma_{\rm
maj}=0.36$ and $\alpha=0.79$ it is   close to an intermediate  galaxy.
NGC~3613 and NGC~4365  are  the only two    core galaxies with   disky
isophotes:  NGC~3613 harbours a thin  edge-on nuclear  disk inside the
central  arcsec, and NGC~4365  contains a kinematically decoupled core
with the characteristics of a cold, rotating system (Surma 1993).

\subsubsection{Central cusp slopes}
\label{sec:cusp-slopes}

In Figure~\ref{fig:gammamag}  we plot $\gamma'_{\rm  maj}$ as function
of both the (major axis) break radius $r_b$  and the absolute $B$-band
magnitude $M_B$.  Solid   circles correspond to core galaxies,   solid
squares   to  power-law galaxies,   and  open circles  to intermediate
galaxies.  All  10 core galaxies have break  radii  in the (relatively
narrow) interval $0.3''  \lta r_b \lta   2''$, and are bright  ($M_B
\lta -20$).  The power-law galaxies, on the other hand, typically have
$M_B  \gta -20$  and $\gamma'_{\rm  maj}  \gta 0.6$.  This once  again
confirms  the dichotomy  amongst   early-type galaxies  inferred  from
previous studies.

But   what  about the    intermediate  galaxies?   According  to their
isophotal structure, one would tend to classify  them as core galaxies
(see  \S~\ref{sec:isoshapes}).  The  same is   true  if one bases  the
classification    on $\gamma$   (cf.     Figure~\ref{fig:gammahisto}).
However,  they also have  properties  that clearly  differentiate them
from core galaxies: their  $\gamma'$-values are more similar  to those
of the power-law galaxies  (although they occupy the low-$\gamma'_{\rm
maj}$ end of the distribution),  their luminosity profiles have values
of $\alpha$  that are significantly lower  than  the typical value for
core galaxies, and finally, their  average absolute magnitudes seem to
be  intermediate  between   those  of  both  the  core  and  power-law
galaxies\footnote{except  for MCG 11-14-25A,  which  is  the  faintest
galaxy  in   our  sample with  $M_B  =    -17.90$}:  they can   not be
unambiguously classified  as  either   one of those.     More detailed
investigations of their properties are required to see if they make up
a truly,   physically distinct  class  of objects,   or whether  these
galaxies  are distinct from both  the core and power-law galaxies only
because of distance and/or resolution effects.

\subsection{Misaligned Structures}
\label{sec:misdisk}

In  fourteen galaxies we   find radial intervals with strong  isophote
twists, non-zero fourth order  moments ($a_4$ and/or $b_4$) and  local
ellipticity maxima.  A clear  example is NGC~2699  where at $\sim 4''$
from the  center  the isophotes   reveal an  isophote  twist  of $\sim
25^{\rm o}$, both $a_4$  and $b_4$ are positive, and  there is a local
maximum in ellipticity.   In some cases the  orientation  angle of the
misaligned structures changes with radius, giving it the appearance of
spiral  arms (e.g.,  ESO~443-39,  NGC~3595, and  UGC~4551),  wheras in
other    cases  (ESO~378-20,   ESO~443-39,  ESO~447-30,  NGC~2950  and
UGC~6062)   misaligned structures are  evident in  two distinct radial
intervals.   A clear example  is ESO~447-30 with misaligned structures
at  both $\sim 2.5 ''$ and  $\sim 10 ''$.   In each of these galaxies,
the two  misaligned structures are not  only misaligned with  the host
galaxy, but also with each other.

Given  the highly  elongated nature of  these structures  and the fact
that   disks are common in  elliptical   galaxies, it  is tempting  to
interpret these structures as stellar disks whose orientation axes are
misaligned   with  those  of  the   host   galaxy.  However,  disks in
spheroidal   systems can  only be  stable  if their  symmetry axis  is
perpendicular to the equatorial plane of the potential, which does not
result  in  a projected  misalignment as  observed.  If, on  the other
hand,  a  stellar disk  is embedded in   a triaxial galaxy, projection
effects can easily accommodate  the observed  misaligned morphologies.
Furthermore, if the axis ratios of the  triaxial host galaxy vary with
radius, the projected angle   of misalignment can vary with  projected
radius.  Such    configuration, therefore, can   explain the  multiple
misaligned  structures observed.   A   problem with this  explanation,
however, is that   all  14 galaxies  with  misaligned   structures are
power-law galaxies  which  are  generally   thought to  be  spheroidal
systems. Furthermore,  if  these  structures  are  disks  the observed
fraction of disks embedded in power-law galaxies is too high (see
\S~\ref{sec:disks} below).

A    more likely  explanation,  therefore, is    that these misaligned
structures correspond to bars. This lends great  support from the fact
that isophotal structures, very similar to  those seen here, have been
observed in barred spirals and  S0s (e.g., Wozniak \etal 1995; Friedli
\etal 1996; Erwin \&  Sparke 1999).  In   particular, the bars  within
bars found   by these authors  provide a  natural  explanation for the
multiple misaligned structures found in some  of our galaxies. Indeed,
both Kormendy (1981) and   Wozniak \etal (1995) have  previously noted
the  complex morphological structure  of NGC~2950 (one of the fourteen
galaxies in our sample with  misaligned structures) and classified  it
as a double-barred galaxy. The bar interpretation is further supported
by the  fact that ten   of the fourteen  galaxies  in our  sample with
misaligned   structures    are  classified   as  S0      or E/S0  (see
Table~\ref{tab:sample}). It is  unclear, however, how to interpret the
misaligned structures in  the four galaxies classified as ellipticals,
although  it is  not    inconceivable  that  these galaxies   may   be
misclassified S0s seen close to face-on.  Detailed investigations into
the kinematics of these  galaxies are required  to confirm that indeed
these structures correspond to tumbling triaxial potentials.

\subsection{Stellar Disks}
\label{sec:disks}

In total 29  out of the 57  galaxies  (51 percent)  in our unperturbed
sample reveal disky isophotes over   at least a small radial  interval
(this     includes    the   misaligned   structures    discussed    in
\S~\ref{sec:misdisk}).  This   diskiness  is   most  straightforwardly
interpreted as due to an embedded stellar disk (e.g., Scorza \& Bender
1995, and references therein).

The diversity of disk structures in our sample  is large, ranging from
extremely thin and small disks in the very  center (i.e., NGC~3613) up
to very large disks spanning the entire  radial interval imaged (i.e.,
NGC~4621).   In 18 galaxies  the isophotes are  disky in only a single
radial interval, whereas the other 11 galaxies have disky isophotes in
two  radial   intervals,   separated by  purely   elliptical  or  boxy
isophotes.  This  ``double-diskiness'' may be  due  to  a nuclear disk
inside the  central hole of  a separate outer disk  or ring (i.e., van
den Bosch \& Emsellem  1998), reflect the projected surface brightness
contribution of a single disk which is locally  dwarfed by that of the
spheroidal,  or   be  due  to    a bar-within-a-bar    structure  (see
\S~\ref{sec:misdisk}).   The    observed isophotal  structure   of  an
elliptical galaxy  with an embedded stellar  disk  depends strongly on
the ratio   of  the  surface  brightness  distributions  of    the two
components and on    the  inclination angle  of   the  disk.  Detailed
disk-bulge decompositions (i.e., Scorza \& Bender  1995; Scorza \& van
den Bosch  1998)  are   required  to decompose the   observed  surface
brightness in the  separate components, which  is beyond the  scope of
this paper.

Given the large fraction  of early-type galaxies that reveal  evidence
for an embedded disk,  and the fact  that these disks are only visible
if the  disk is seen  under a  sufficiently  large inclination  angle,
implies that the true   fraction of early-type galaxies with  embedded
stellar disks  is likely   to  be significantly larger than   the $51$
percent found here.  In our unperturbed  sample, $66$ percent of all
power-law   galaxies have some    amount   of diskiness.  This   is  a
surprisingly high fraction.  If we  assume that all power-law galaxies
have embedded disks, and that these disks are only detectable if their
projected minor-to-major  axis ratio is smaller  than $0.5$,  one only
expects to detect disks in $33$ percent of all power-law galaxies (see
Rix \& White  1990 for a detailed discussion  on  the detectability of
embedded   stellar disks).   The observed fraction   of $66$ percent
implies either of  the following: (i) all disks  with a projected axis
ratio smaller than $0.86$  are detectable (assuming that all power-law
galaxies  harbor embedded disks),  (ii)  power-law galaxies are always
seen under high inclination angles, (iii) the  steep central cusp owes
to  the projection of the   highly flattened structure (but see  Lauer
\etal 1995; Faber \etal 1997), or (iv) in  some cases the diskiness is
due to highly elongated   prolate or triaxial structures  (i.e., bars)
rather than oblate disks.

We only   consider the latter  of these  explanations viable.    It is
supported by the  fact that fourteen of   our power law galaxies  have
their diskiness  misaligned with the main galaxy.   As we discussed in
the previous section, these structures are  most likely due to bars in
S0 galaxies.  If we exclude these fourteen  galaxies, we find embedded
disks  in only $32$  percent of our power-law  galaxies, which is in
excellent agreement  with the $33$  percent expected if  all power-law
galaxies  harbor embedded disks, but which  are only detectable if the
disk's projected axis ratio is smaller than $0.5$.

\subsection{Nucleated Galaxies}
\label{sec:nuclei}

Although the Nuker-law in general provides a  good fit to the observed
surface brightness of the central  regions of early-type galaxies,  in
several cases the very centers reveal excess light, associated with an
increase in slope as the HST resolution limit is approached. We follow
Lauer \etal (1995) and refer to these features as {\it nuclei}.  These
unresolved nuclei may be due to non-thermal emission  from AGNs or may
reflect the presence of a nuclear star cluster.

Table~\ref{tab:isoparameter}  lists the degree of nucleation observed,
ranging from  very mild, indicated by `I'  (e.g.,  NGC~2634) to nuclei
that stand out strongly  against  the `background' Nuker-law  profile,
indicated  by  `III'  (e.g.,    NGC~4482).  In  order  to  be   termed
`nucleated'  we require  that  the central steepening  of the  surface
brightness occurs inside $0.15''$, and  is visible from both the major
and minor axes surface brightness  profiles.  Note that effects due to
deconvolution as well as dust obscuration can produce artefacts in the
luminosity profiles that  are similar to our  level I  nucleation.  In
order  to guard against  such  artificial  `detections', we have  been
fairly conservative in assigning nucleation.

In total, nuclei are detected in 9 out of the 67 galaxies (13 percent)
in  our sample.  This  is less than  half the  detection rate of Lauer
\etal (1995), who detected nuclei in 16 out of 45 galaxies. Except for
the fact that we have been fairly conservative in assigning nuclei, it
is unclear why our detection rate is so much smaller. Only 2 of the 16
nucleated galaxies in the sample of Lauer \etal are core galaxies.  In
both cases the nucleation is interpreted  as non-thermal emission from
AGNs.  In   our sample, the   two  nucleated galaxies   (NGC~4168  and
NGC~5077) not classified as  power-law galaxies have been detected  at
1.4 GHz (see Table~\ref{tab:sample}). This is consistent with the idea
that `nucleation' in   core (and intermediate)  galaxies is associated
with an   active galactic  nucleus.    For  none  of the  7  nucleated
power-law   galaxies  do we   find any sign   of  non-thermal emission
associated with  an  AGN (i.e.,  none  of   these galaxies have   been
detected  at  1.4  GHz).  This again   is  in good  agreement with the
results  of Lauer \etal (1995).  These  results, however, are somewhat
puzzling from  an  astrophysical point of  view.   An often advertised
interpretation  of (non-AGN related)  nuclei is that  they are stellar
remnants of cannibalized,  low-mass  galaxies  that have reached   the
center by means  of dynamical friction.   Since less luminous galaxies
are  in   general denser than  their   more  massive counterparts, one
expects to  find stellar nuclei predominantly  in the less dense cores
of the   more massive galaxies, where   tidal forces  are  less severe
(e.g., Kormendy 1984; Faber \etal 1997).   The almost complete absence
of stellar nuclei  in core galaxies hints  at  some violent disruption
process, most likely associated  with massive black holes (see Merritt
\& Cruz 2001, and references therein).  However, this fails to explain
why  power-law galaxies   do  seem to harbor    stellar nuclei.  These
galaxies have  massive black  holes as  well,  and their intrinsically
high stellar densities will only  aid in disrupting any merger remnant
that might reach the  center  of the  potential well.  A  more  viable
explanation is that these nuclei are the product of some dissipational
formation mechanism.  The presence of highly flattened disk-structures
in  the majority of  these  power-law galaxies  supports the view that
dissipational processes have played an important role.

\section{Conclusions}
\label{sec:concl}

We have reported  on the $R$-band images  of the central regions of 67
early-type  galaxies  obtained with  the  WFPC2 aboard the  HST.  This
sample roughly doubles the number of early-type galaxies that have now
been imaged at  HST resolution,  and  complements similar  data on the
central regions of radio  galaxies and the  bulges of spiral galaxies. 
In  this paper we have presented  the data, focusing  in particular on
the isophotal structures and luminosity profiles.

Our sample strikingly confirms the complex morphologies of the stellar
component in the  central regions of   early-type galaxies which  have
become  apparent  from previous studies  with  HST.  Whereas the outer
regions of early-type galaxies can in  general be classified as either
disky or boxy, the isophotal structure in the inner $\sim 10''$ varies
strongly with radius,  such   that a bimodal  classification  does not
suffice to  describe  the  bewildering variety  of structures   seen.  
Instead, we have used  a simple letter  coding to describe the diverse
morphologies.

Previous  imaging  surveys with HST  have  shown that a  more suitable
criterion  for classifying the central  regions of early-type galaxies
is based  on  their luminosity  profiles  (rather than their isophotal
structure).  Depending on the slope of  the central cusp, galaxies are
classified as   either   a core  galaxy or  a   power-law galaxy.   We
determined the luminosity profiles along the major  and minor axes for
57 galaxies in our sample  (the unperturbed sample).  The remaining 10
galaxies do not allow a proper assessment of their luminosity profiles
because of dust  obscuration. We have   shown how luminosity  profiles
derived from an isophotal analysis of  the data can produce relatively
large systematic errors in the central region, owing to  the sub-pixel
interpolation  used.
We therefore use  the actual pixel  values in  the  central $0.2''$ to
define the luminosity profiles.  We fit the luminosity profiles with a
Nuker-law profile (equation~[\ref{nukerlaw}]), which has been shown to
adequately describe the  luminosity profiles of early-type galaxies in
the  inner  $\sim 20''$.  At  radii   less than $0.2''$   we fit  this
Nuker-law model directly to the observed pixel values, while at larger
radii we  use the luminosity profile  as determined from the isophotal
analysis.  Detailed modeling has   shown  that this  method allows  an
accurate recovery of the actual luminosity profiles.

In a similar study, based on a sample of 61 early-type galaxies imaged
with   the WFPC1,   Faber~\etal  (1997)  classified   galaxies as core
galaxies if the central cusp slope $\gamma$ of  the best fitting Nuker
law was less than $0.3$.  Galaxies with $\gamma > 0.3$ were classified
as power-law galaxies.   In our sample   we find a subset of  galaxies
with luminosity profiles  that have   a continuously changing   slope,
rather  than a double power-law shape  with a  clear break radius. For
these  systems $\gamma$ of the   best-fit Nuker-law profile is smaller
than $0.3$, but it is only a measure of the gradient of the luminosity
profile at radii much smaller than the resolution  limit of the actual
observations.  At    the resolution  limit the  actual    slope of the
luminosity profile is steeper than $0.3$, and it is unclear whether to
classify these  systems as core or  power-law galaxies.   Although the
isophotal properties are similar to those  of the core galaxies, there
is a   hint that the  luminosities of  these galaxies are intermediate
between  those  of  the core and    power-law galaxies.   We therefore
provisionally classify  these galaxies as `intermediate'.  However, it
is unclear  at  present whether these  galaxies make  up  a physically
distinct class  or  whether distance  and/or resolution effects  cause
them to loose their distinct core and/or power-law characteristics.

F97 have pointed  out that the distribution of  central cusp slopes is
clearly bimodal: whereas (by definition)  core galaxies have $\gamma <
0.3$, power-law galaxies  are  found to have  $\gamma  \gta  0.5$. The
region $0.3  < \gamma < 0.5$  is  virtually devoid  of galaxies.  This
bimodality closely follows the dichotomy in other properties: the core
galaxies  make  up the  bright  end  of  the  luminosity function  and
typically  have  boxy isophotes,  whereas   the power-law galaxies are
fainter and typically have disky isophotes.  We confirm this dichotomy
with   our sample:  93  percent   of all  galaxies  that reveal  disky
isophotes  are power-law galaxies, whereas   77  percent of all   core
galaxies have    either    boxy  or  purely   elliptical    isophotes.
Furthermore, we confirm the bimodality of  cusp slopes, but we do find
a  significant population of  galaxies   with $0.3  < \gamma <   0.5$.
Combined  with the class of   intermediate  galaxies, which have  cusp
slopes at the  resolution limit that also fall  in this range, we thus
conclude that the bimodality is not as strong as previously claimed.

Finally, the statistical properties of our sample can be summarized as
follows:
\begin{itemize}

\item In 15 percent of the galaxies in our  sample we detect dust
  disks with diameters that range from $0.3''$  to $12.3''$.  Together
  with the  28 percent of the galaxies  in which we detect filamentary
  dust, the  total  detection rate  of dust  is  43  percent, in  good
  agreement with previous estimates  based on HST  data (van Dokkum \&
  Franx 1995).

\item In 9 galaxies we find evidence for nuclei. These are defined as
  (largely) unresolved, nuclear surface brightness spikes that cause a
  central  upturn  in the luminosity  profiles.  Our detection rate of
  nuclei (13 percent) is substantially  less than the 36 percent found
  by Lauer \etal  (1995) in their sample, but   this is most likely  a
  reflection  of   our  more   conservative  approach to   identifying
  nucleation. Whereas the nuclei in power-law galaxies are most likely
  stellar,   the  nucleation seen    in   core galaxies  is  generally
  non-thermal emission from an active galactic nucleus.

\item  In 14 galaxies we find  structures that are misaligned with the
  principal axes of the host galaxy.  These structures are most likely
  associated with stellar bars in S0 galaxies. In 5 of these galaxies,
  multiple misaligned  structures are apparent  indicating bars within
  bars.

\item A large fraction (51 percent) of  the early-type galaxies in our
  unperturbed  sample reveal  disky isophotes  over at   least a small
  radial    interval, whereas this   fraction   increases to almost 70
  percent when only   considering  power-law  galaxies.  This  is    a
  remarkably large  fraction: even if all  power-law  galaxies were to
  harbor  an embedded stellar  disk, one  would  not expect  to detect
  these thin  structures in more than $\sim  30$ percent  of all cases
  because of inclination effects.   If, however, we exclude the  cases
  where  the diskiness is misaligned  with the main  galaxy, and which
  are more likely associated with stellar bars, we find that $\sim 32$
  percent of  all power-law galaxies reveal some  amount of diskiness. 
  This  is consistent with  a   picture in  which {\it  all} power-law
  galaxies harbor an embedded stellar disk.

\end{itemize}


\acknowledgments  

We are grateful to  the  referee, Tod  R.  Lauer, for  his  insightful
comments that greatly improved  the  analysis and presentation of  the
data  in this  paper.  The observations  presented  in this paper were
obtained with the  NASA/ESA Hubble Space  Telescope, which is operated
by AURA, Inc., under NASA contract NAS 5-26555. We acknowledge the use
of the LEDA  database (www-obs.univ-lyon1.fr).  FvdB was supported  by
NASA through  Hubble  Fellowship grant \#  HF-01102.11-97.A awarded by
the Space Telescope Science Institute.


\clearpage

\begin{appendix}

\section{Contour maps, luminosity profiles, and isophotal parameters}
\label{sec:App}

In the following  figures we present the  data for all  67 galaxies in
our sample.  The  upper right-hand panels   show contour plots of  the
$33.0'' \times 33.0''$ field  of view  of the  Planetary  Camera.  The
thick black bar in the  upper-right corner  corresponds to an  angular
size  of  $5.0''$.  Contours are   plotted at  intervals of  $0.3$ mag
between $\mu_R = 22.5$ mag arcsec$^{-2}$ and the surface brightness at
$r = 1.0''$ (in order to prevent central crowding of contours).

The upper-panels  on the  left show  the observed luminosity  profiles
(open  circles are determined  from   the isophotal analysis,   closed
circles are the actual pixel values fitted and plotted as described in
\S~\ref{sec:models}) along the major  (upper curves) and  minor axis 
(lower  curves, offset by  $+0.5$ mag  for  clarity).  The thick solid
lines are the best fit Nuker-law profiles, the parameters of which are
listed  in Tables~\ref{tab:NukerMajor}  and~\ref{tab:NukerMinor}.   In
some cases data in the central regions  is excluded from the Nuker-law
profile fit because of nucleation and/or nuclear dust disks.  In these
cases a vertical  thin  dashed line indicates the  lower  limit of the
data range used.

The   remaining panels  plot the   isophotal parameters  for both  the
non-deconvolved image   (solid  triangles) and the  deconvolved  image
(open circles with errorbars). For each galaxy  we plot as function of
radius  the  ellipticities ($\epsilon$), the   position  angles of the
isophotes  ($\theta$, measured   from    North  to  East),  and    the
higher-order parameters $a_n$ and $b_n$ (with $n=3,4$). 

Some galaxies  have  their central  regions  saturated in the  700 sec
exposure. In  these cases we plot  the isophotal parameters of the 300
sec exposures in  the radial regime  where the 700 sec  exposure shows
saturation  effects (rather  than those  of the  combined  image).  In
three cases (NGC~2950, NGC~3377, and NGC~4621), the central regions of
the 300 sec exposures are also saturated. For these galaxies, only the
isophotal parameters are plotted  down to the radius where  saturation
effects in the 300 sec exposure start.

\end{appendix}

\clearpage
 

\ifsubmode\else
  \baselineskip=10pt
\fi


\clearpage


\ifsubmode\else
\baselineskip=14pt
\fi


\submodefigcaption{\figmodeliso}{Isophotal parameters as a function
  of   radius  for a  model   galaxy  which has  perfectly  elliptical
  isophotes  with  $\epsilon = 0.26$    independent  of radius, and  a
  central       cusp    with   $I(r)      \propto    r^{-0.7}$    (cf.
  Figure~\ref{fig:modelsb}).    Solid    triangles correspond  to  the
  parameters  of the (unconvolved)  raw image,  in which only  Poisson
  noise is introduced.  The open circles with error bars correspond to
  the parameters   of the deconvolved  image.   Note that for  $r \gta
  0.2''$ the parameters   of the  raw  and deconvolved  images  are in
  excellent  agreement.  We therefore are confident  that we can trust
  the isophotal  parameters of our  deconvolved  images at  least  for
  semi-major axis radii larger than  $\sim 0.2''$. For $r \lta 0.2''$,
  however, the    sub-pixel  interpolation   of   the  fitting-routine
  systematically underestimates  the ellipticity of the isophotes, and
  causes  artificial  deviatations   from  their  perfectly elliptical
  shapes. \label{fig:modeliso}}

\submodefigcaption{\figdeltamu}{Characterization of errors in
  luminosity profiles.  For each   model $i$, the  difference  $\Delta
  \mu_{i}(r)$  between  the input  surface   brightness profile (i.e.,
  equation~[\ref{deltamu}])  and   the  major axis  luminosity profile
  determined from  the isophotal analysis  of the unconvolved image is
  calculated.  For a sample of  8 sets of model  galaxies (each set is
  characterized by the same   value of $\gamma$) the average  $\langle
  \Delta \mu \rangle(r)$ (right  panel) and the corresponding standard
  deviation   $\sigma_{\Delta\mu}(r)$  (left panel) is  then computed,
  where the average is over the 50 models  per set.  The number in the
  legend corresponds to the characteristic central cusp slope $\gamma$
  of  the specific set.  For   $r \gta 0.2''$ the luminosity  profiles
  derived from the isophotal analysis  are in excellent agreement with
  the  intrinsic  profile.  For   smaller  radii, however,  the  error
  increases strongly, especially for the  profiles with steep  central
  cusp slopes.  Note  also  that the average residual  $\langle \Delta
  \mu \rangle(r)$ strongly deviates from zero at small radii, implying
  that the main contribution to the error is of systematic rather than
  random nature.   This  is  a reflection of   the problems  with  the
  isophotal analysis due to the discrete, pixelized nature of the data
  and the sub-pixel interpolation.\label{fig:deltamu}}

\submodefigcaption{\figmodelsb}{In the upper panels, the solid lines
  correspond to  the  intrinsic major axis  luminosity  profile of the
  same model  as in Figure~\ref{fig:modeliso}.  Upper profiles (filled
  circles) correspond to  the luminosity profiles determined  from the
  isophotal analysis. Lower  profiles (open  symbols),  offset by  1.5
  magn,  correspond to the  luminosity profile determined with the new
  method utilizing the actual  pixel values inside  $ r < 0.2''$ (open
  squares) combined with the isophotal analysis data at $r \geq 0.2''$
  (open circles). The lower  panels display the  residuals for the two
  differents methods (same symbols as in upper panels). As is evident,
  the new method allows us to recover the intrinsic luminosity profile
  with high accuracy down   to  very small radii, whereas   luminosity
  profiles determined solely  from the  isophotal analysis result   in
  relatively large errors at small radii.\label{fig:modelsb}}

\submodefigcaption{\figfchi}{The left panel plots the average
  difference between the intrinsic cusp steepness $\gamma_{\rm model}$
  and the best-fit Nuker-law profile $\gamma_{\rm fit}$ for the set of
  models with $\gamma=0.5$ as a function of $f$.  A value of $f \simeq
  5$  yields  values for $\gamma_{\rm  fit}$   in good agreement  with
  $\gamma_{\rm model}$.  For $f \lta  1$, too much relative weight  is
  given to the pixel values inside $0.2''$,  and $\gamma_{\rm fit}$ is
  too large.  For too large $f$,  $\chi^2$  is completely dominated by
  the  isophotal  luminosities outside $0.2''$    and the central cusp
  slope  becomes poorly  constrained,  as is   evident from the  large
  errorbars. We therefore adhere to  $f=5$ for the final analysis. The
  right panel plots the  fitted central cusp slope $\gamma_{\rm fit}$,
  determined  with  the  new    fitting    method as   described    in
  \S~\ref{sec:models}, versus the intrinsic   slope. The  dots,  which
  correspond to a  single model, are grouped  in 15 sets  of 50 models
  each.  Note that $\gamma_{\rm    fit}$  is in  good agreement   with
  $\gamma_{\rm model}$.\label{fig:fchi}}

\submodefigcaption{\figgamma}{Correlations between the two different
  measures of the central cusp  of the luminosity profile: $\gamma$ is
  the asymptotic logarithmic slope  of the best-fit Nuker-law profile,
  and $\gamma'$  is the logarithmic   slope of that  fit at $r=0.1''$.
  The indices   `maj' and `min' refer to   the major  and  minor axes,
  respectively.  Both $\gamma$ and $\gamma'$ give very similar results
  along both axes as shown in the upper  panels. Comparing $\gamma$ to
  $\gamma'$ yields a  more complicated pixture.  Galaxies with $\gamma
  > 0.5$ in general have $\gamma' \simeq \gamma$. However, for $\gamma
  < 0.3$  about  half of the  galaxies  have $\gamma' <  0.3$, and the
  other half has  $\gamma' > 0.3$.  See  the text for  a more detailed
  discussion. \label{fig:gamma}}

\submodefigcaption{\figgammahisto}{Histograms of $\gamma$ (left
  panels)    and  $\gamma'$ (right panels)  for   the  galaxies in our
  unperturbed sample.  The indices `maj' and `min'  refer to the major
  and  minor  axes,  respectively.   Black  areas correspond  to  core
  galaxies,   cross-hatched areas    to    power-law   galaxies,   and
  single-hatched  areas to  the intermediate   galaxies.  Although the
  distribution of cusp slopes reveals a clear  hint for bimodality, we
  find less of a deficit of  galaxies with $0.3  \leq \gamma \lta 0.5$
  than F97. Note also that the intermediate galaxies themselves follow
  the same distribution  of $\gamma$ as the core  galaxies, but at the
  same time   occupy    the  low-$\gamma'$   end   of  the   power-law
  galaxies. \label{fig:gammahisto}}

\submodefigcaption{\figgammamag}{The parameter $\gamma'_{\rm maj}$  as
  function  of break radius $r_b$ (left   panel) and absolute $B$-band
  magnitude $M_B$ (right panel).  Core galaxies are indicated by solid
  circles,   intermediate  galaxies by   open  circles,  and power-law
  galaxies by solid squares.  Note that the intermediate galaxies seem
  to have properties intermediate between those  of core and power-law
  galaxies. \label{fig:gammamag}}

\setcounter{figure}{0}

\dofig{\figmodeliso}  {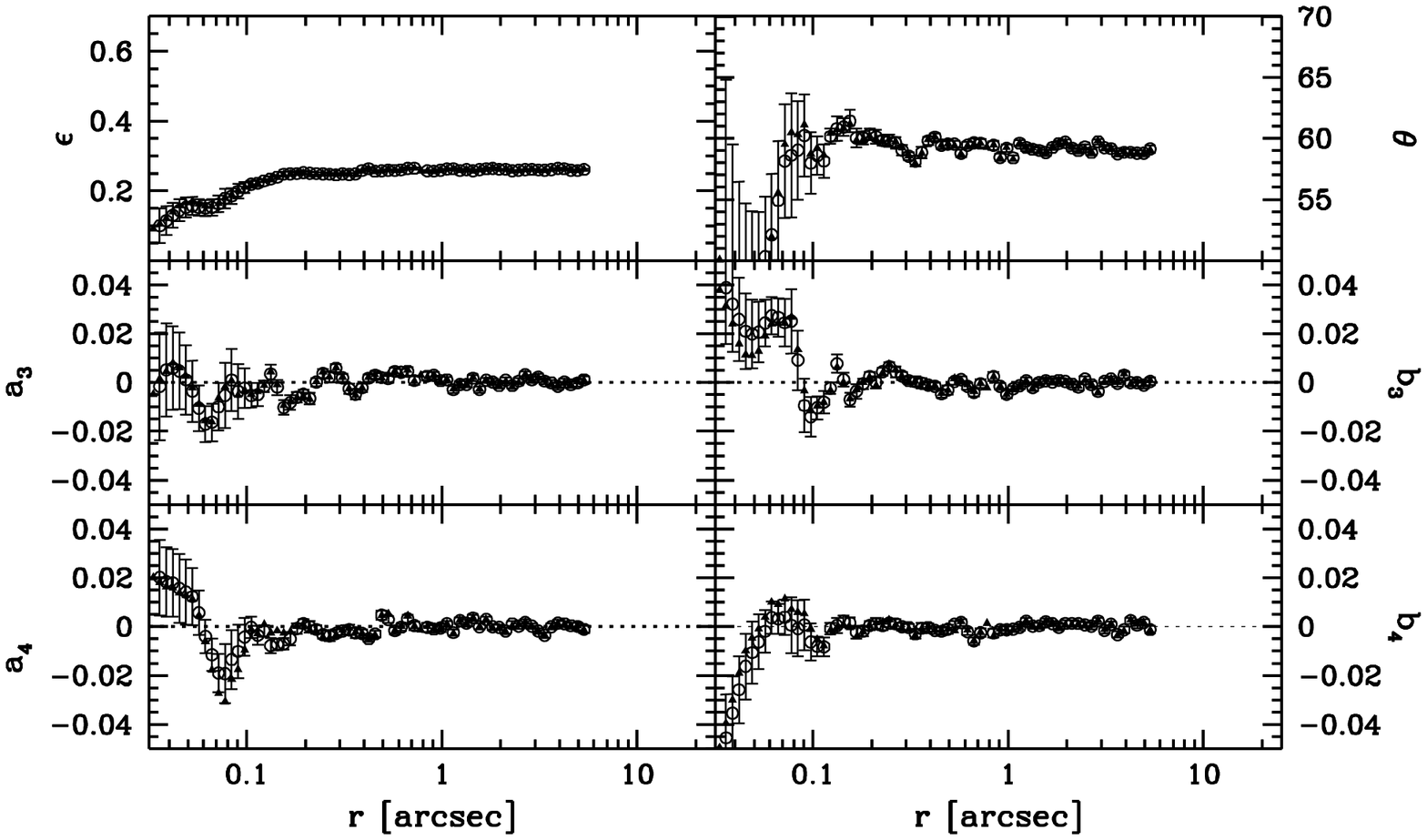}{16.0}{0.0}
\dofig{\figdeltamu}   {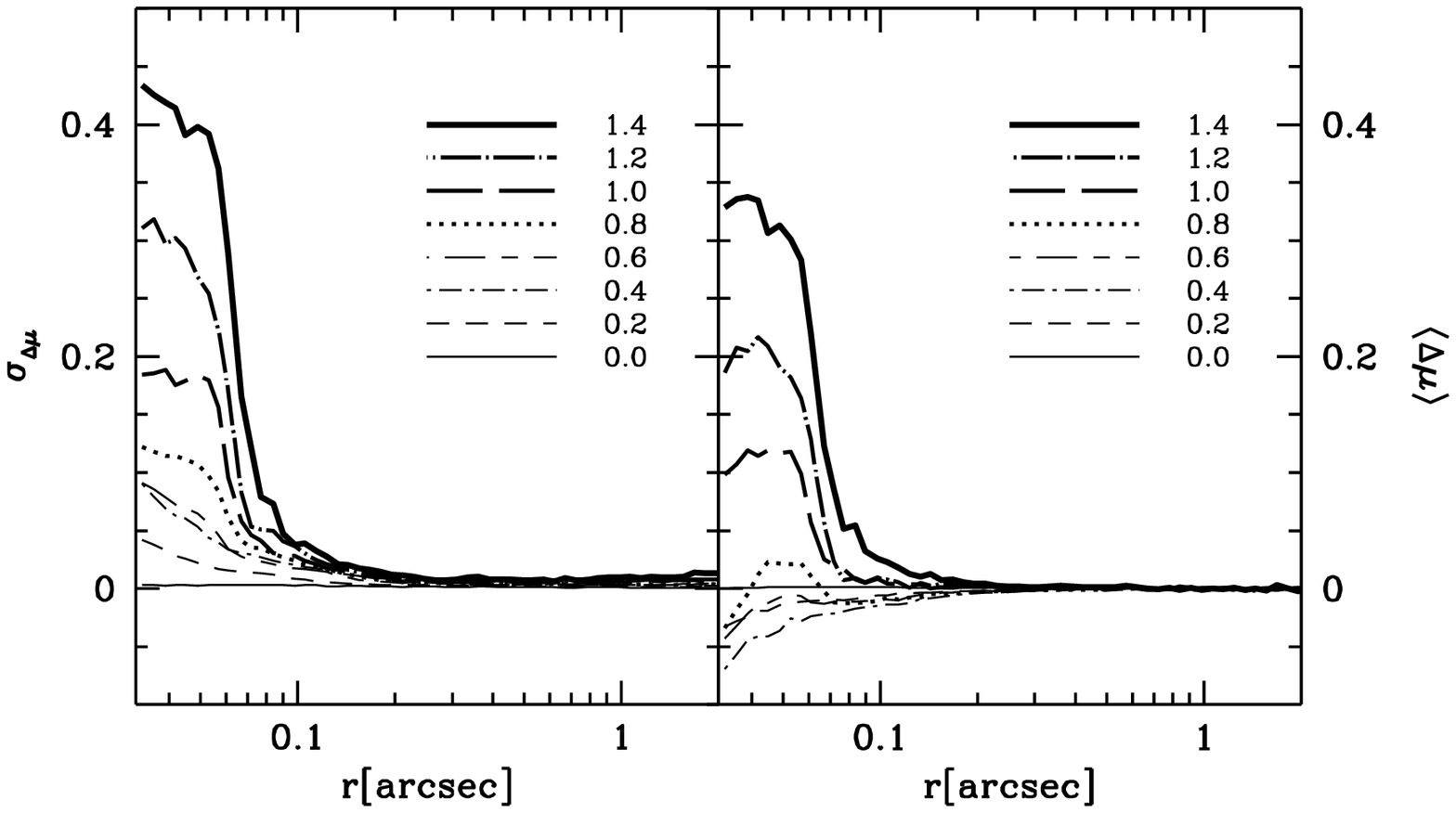}{16.0}{0.0}
\dofig{\figmodelsb}   {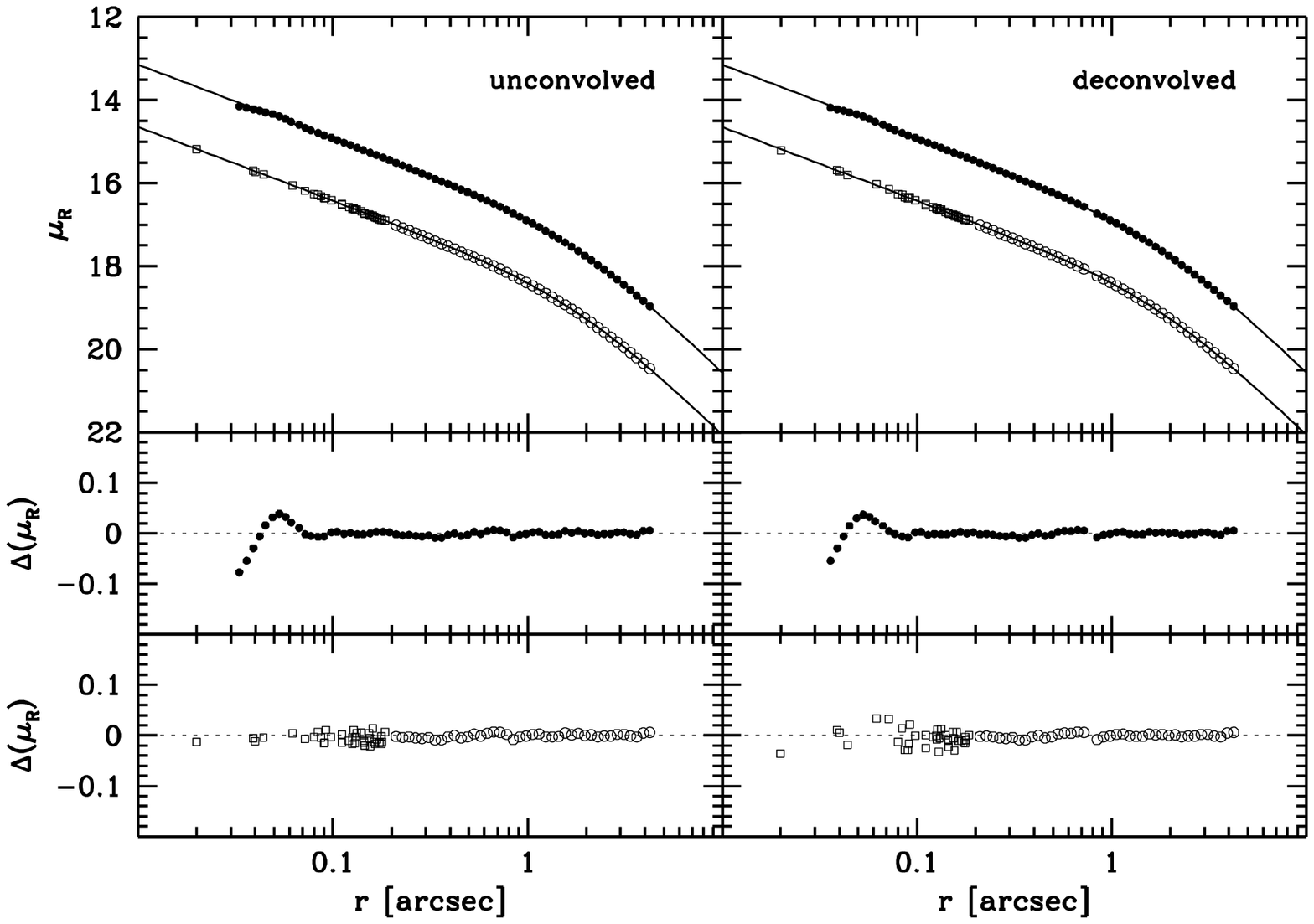}{16.0}{0.0}
\dofig{\figfchi}      {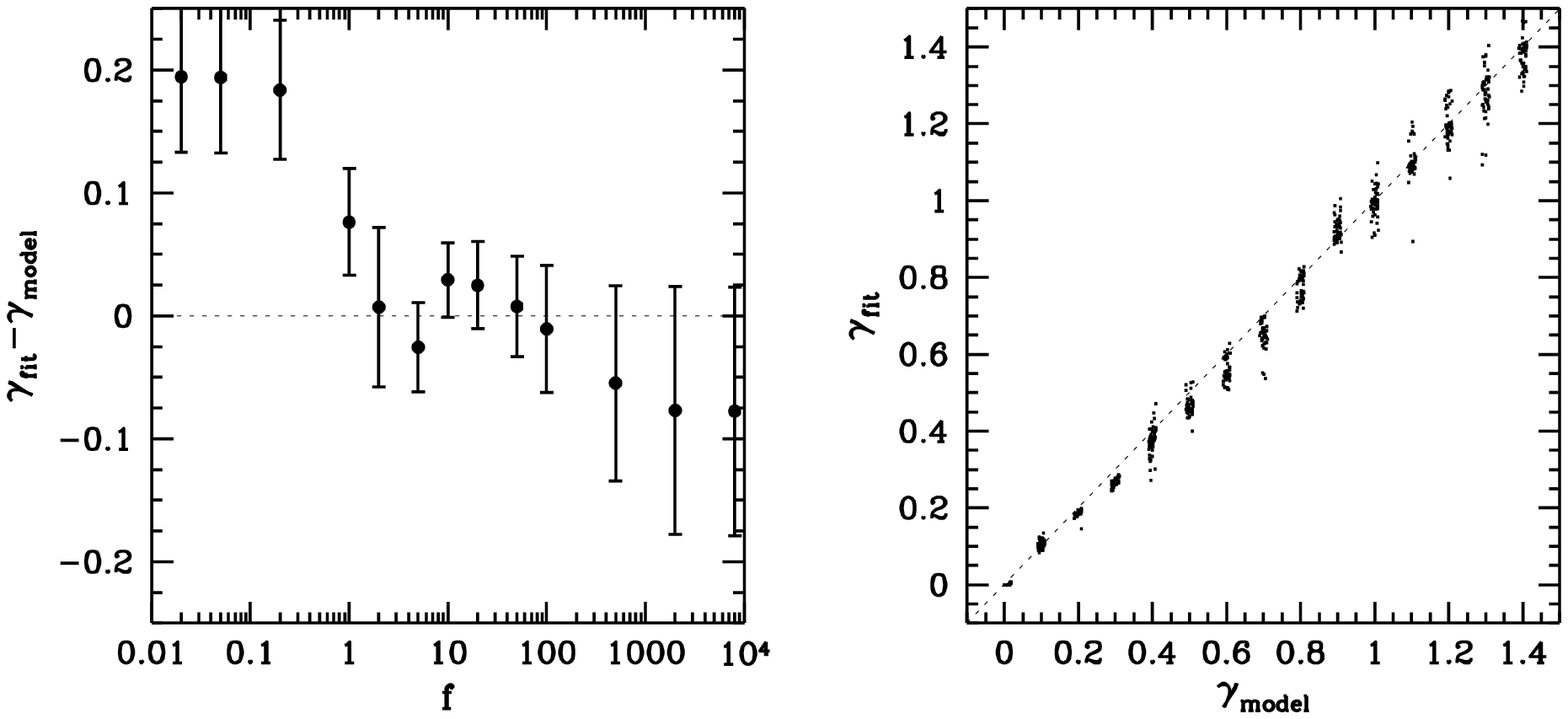}{16.0}{0.0}
\dofig{\figgamma}     {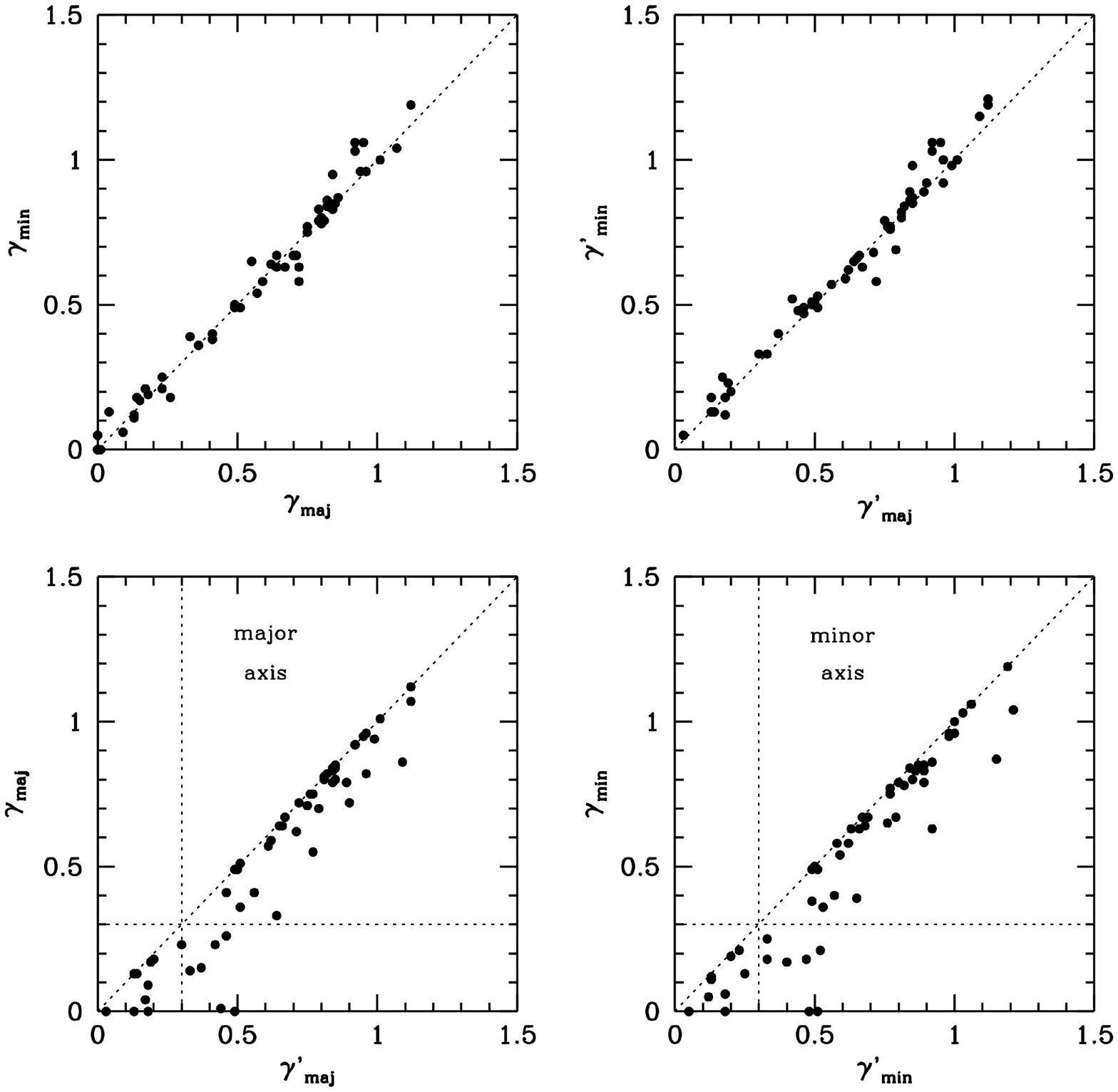}{16.0}{0.0}
\dofig{\figgammahisto}{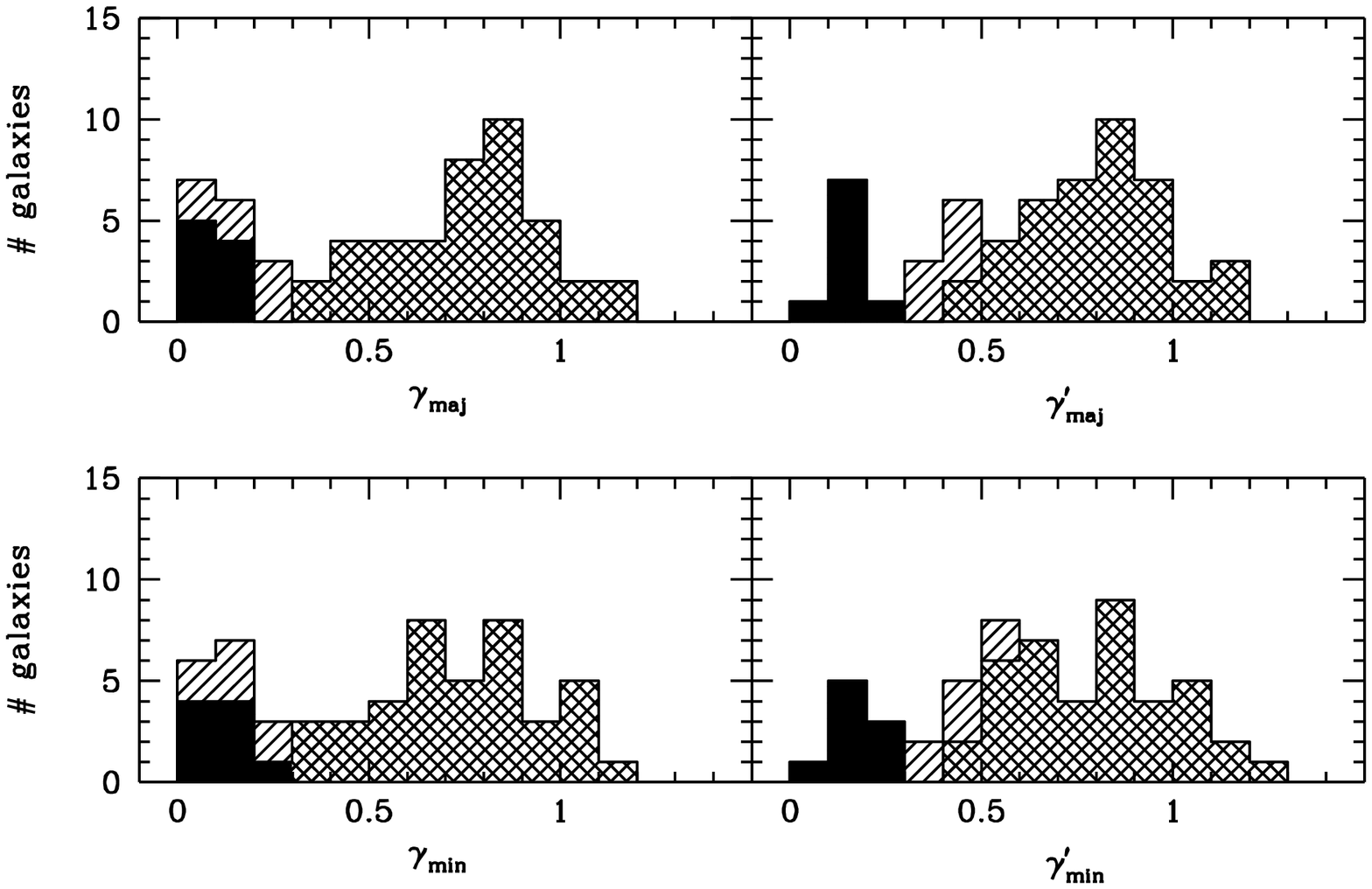}{16.0}{0.0}
\dofig{\figgammamag}  {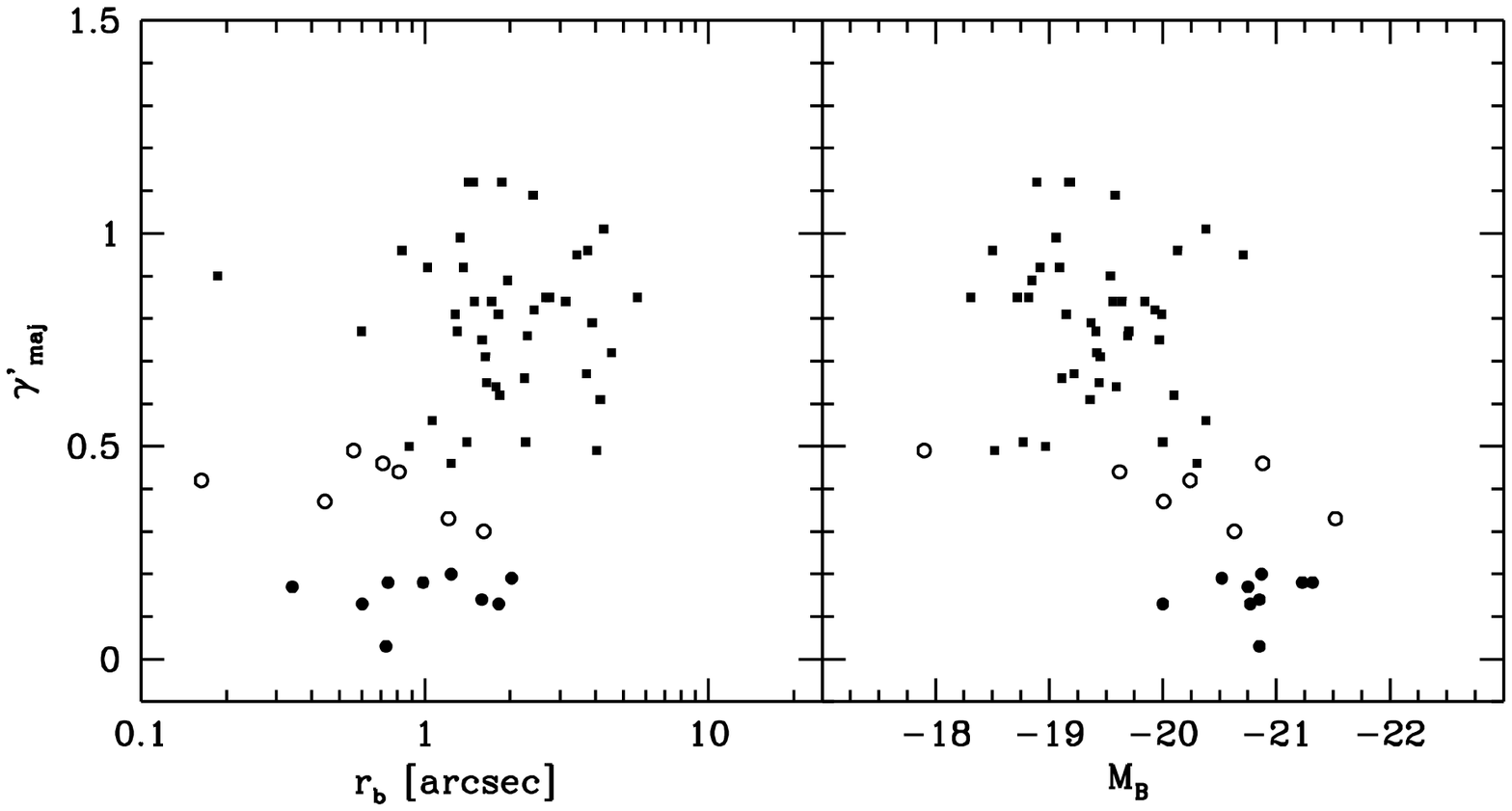}{16.0}{0.0}

\ifsubmode
  \clearpage
\fi



\clearpage
\ifsubmode\pagestyle{empty}\fi


\begin{deluxetable}{lcccccccccc}
\tablewidth{0pt}
\tablecaption{Global parameters.\label{tab:sample}}
\tablehead{\colhead{Name} & \colhead{$M_B$} &\colhead{$B_T$} & 
\colhead{$D$} & \colhead{Type} & \colhead{$\epsilon_{25}$} & 
\colhead{$d_{25}$} & \colhead{$V_{\rm vir}$} & \colhead{$B-V$} &
\colhead{$f_{1.4}$} & \colhead{$m_{\rm FIR}$} \\
\colhead{(1)} & \colhead{(2)} & \colhead{(3)} & \colhead{(4)} & 
\colhead{(5)} & \colhead{(6)} & \colhead{(7)} & \colhead{(8)} &
\colhead{(9)} & \colhead{(10)} & \colhead{(11)} \\
}
\startdata
ESO 378-20    & $-19.58$ & $13.17$ & $ 35.6$ & S0a  & $0.53$ & $1.134$ & $3032$ & $0.756$ &               & \nl        
ESO 437-15    & $-19.41$ & $13.14$ & $ 32.3$ & S0   & $0.78$ & $1.370$ & $2753$ &         & $3.5 \pm 0.6$ & $13.46$ \nl        
ESO 443-39    & $-19.41$ & $13.39$ & $ 36.3$ & S0   & $0.43$ & $1.146$ & $3042$ &         &               & \nl        
ESO 447-30    & $-19.84$ & $12.88$ & $ 35.0$ & S0   & $0.26$ & $1.233$ & $2917$ &         &               & $13.88$ \nl        
ESO 507-27    & $-19.37$ & $13.56$ & $ 38.5$ & S0   & $0.76$ & $1.238$ & $3202$ &         &               & \nl       
ESO 580-26    & $-18.12$ & $14.85$ & $ 39.3$ & S0   & $0.23$ & $0.998$ & $3215$ &         & $3.7 \pm 0.7$ & $13.15$ \nl
IC 875        & $-19.17$ & $13.73$ & $ 37.9$ & S0   & $0.18$ & $1.184$ & $2794$ & $0.731$ &               & \nl        
MCG 11-14-25A & $-17.90$ & $15.32$ & $ 44.2$ & E/S0 & $0.12$ & $0.752$ & $3304$ &         &               & \nl        
MCG 8-27-18   & $-18.85$ & $14.36$ & $ 43.7$ & E/S0 & $0.05$ & $1.002$ & $3276$ &         &               & \nl        
NGC 2549      & $-19.22$ & $11.76$ & $ 15.7$ & S0   & $0.68$ & $1.600$ & $1064$ & $0.842$ &               & \nl        
NGC 2592      & $-18.92$ & $13.11$ & $ 25.5$ & E    & $0.16$ & $1.273$ & $1987$ &         &               & \nl        
NGC 2634      & $-19.99$ & $12.50$ & $ 31.4$ & E    & $0.00$ & $1.358$ & $2268$ & $0.823$ &               & \nl        
NGC 2699      & $-18.72$ & $12.97$ & $ 21.8$ & E    & $0.06$ & $1.128$ & $1825$ &         &               & \nl        
NGC 2778      & $-19.06$ & $13.06$ & $ 26.6$ & E    & $0.24$ & $1.147$ & $2024$ & $0.850$ &               & \nl        
NGC 2824      & $-19.75$ & $13.92$ & $ 54.2$ & S0   & $0.36$ & $0.998$ & $4268$ & $0.784$ & $9.3 \pm 0.5$ & $13.26$ \nl 
NGC 2872      & $-20.38$ & $12.54$ & $ 38.3$ & E    & $0.25$ & $1.279$ & $3070$ & $0.904$ & $7.4 \pm 2.6$ & $12.09$ \nl        
NGC 2902      & $-18.97$ & $12.86$ & $ 23.3$ & S0   & $0.22$ & $1.256$ & $1991$ &         &               & \nl        
NGC 2950      & $-19.93$ & $11.52$ & $ 19.5$ & S0   & $0.32$ & $1.487$ & $1353$ & $0.811$ &               & \nl        
NGC 2986      & $-20.87$ & $11.27$ & $ 26.8$ & E    & $0.12$ & $1.588$ & $2302$ & $0.861$ &               & \nl        
NGC 3065      & $-19.56$ & $12.69$ & $ 28.2$ & S0   & $0.05$ & $1.300$ & $2009$ & $0.873$ & $4.5 \pm 0.6$ & \nl        
NGC 3078      & $-20.71$ & $11.60$ & $ 29.0$ & E    & $0.17$ & $1.512$ & $2491$ & $0.890$ & $314 \pm  11$ & \nl        
NGC 3193      & $-19.62$ & $11.64$ & $ 17.9$ & E    & $0.08$ & $1.315$ & $1373$ & $0.885$ &               & \nl        
NGC 3226      & $-19.11$ & $12.08$ & $ 17.3$ & E    & $0.12$ & $1.448$ & $1337$ & $0.835$ & $3.3 \pm 0.5$ & $10.86$ \nl        
NGC 3266      & $-19.11$ & $13.18$ & $ 28.6$ & S0   & $0.12$ & $1.166$ & $2057$ &         &               & \nl        
NGC 3348      & $-21.23$ & $11.69$ & $ 38.5$ & E    & $0.00$ & $1.354$ & $2826$ & $0.918$ & $8.3 \pm 0.5$ & \nl        
NGC 3377      & $-18.89$ & $10.90$ & $  9.1$ & E    & $0.39$ & $1.663$ & $ 698$ & $0.801$ &               & \nl        
NGC 3414      & $-19.64$ & $11.73$ & $ 18.8$ & S0   & $0.17$ & $1.557$ & $1414$ & $0.898$ & $4.7 \pm 0.5$ & $14.97$ \nl        
NGC 3595      & $-19.69$ & $12.72$ & $ 30.4$ & E/S0 & $0.53$ & $1.225$ & $2248$ &         &               & \nl        
NGC 3613      & $-20.75$ & $11.49$ & $ 28.0$ & E    & $0.45$ & $1.576$ & $2021$ & $0.849$ &               & \nl        
\tablebreak
NGC 3640      & $-20.01$ & $11.03$ & $ 16.2$ & E    & $0.10$ & $1.684$ & $1312$ & $0.853$ &               & \nl        
NGC 4121      & $-18.31$ & $13.95$ & $ 28.3$ & E    & $0.12$ & $0.645$ & $2024$ & $0.824$ &               & \nl        
NGC 4125      & $-21.20$ & $10.31$ & $ 20.1$ & E    & $0.16$ & $1.822$ & $1361$ & $0.850$ &               & $14.24$ \nl
NGC 4128      & $-19.97$ & $12.58$ & $ 32.4$ & S0   & $0.65$ & $1.400$ & $2339$ & $0.820$ &               & \nl        
NGC 4168      & $-20.52$ & $11.83$ & $ 29.5$ & E    & $0.18$ & $1.468$ & $2310$ & $0.852$ & $6.0 \pm 1.5$ & \nl        
NGC 4233      & $-19.71$ & $12.64$ & $ 29.6$ & S0   & $0.53$ & $1.374$ & $2340$ & $0.882$ & $3.4 \pm 0.6$ & \nl        
NGC 4291      & $-20.00$ & $12.04$ & $ 25.6$ & E    & $0.14$ & $1.350$ & $1781$ & $0.865$ &               & \nl        
NGC 4365      & $-20.77$ & $10.21$ & $ 15.7$ & E    & $0.24$ & $1.834$ & $1232$ & $0.892$ &               & \nl        
NGC 4474      & $-19.42$ & $12.13$ & $ 20.4$ & S0   & $0.35$ & $1.404$ & $1571$ & $0.788$ &               & \nl        
NGC 4478      & $-19.36$ & $11.93$ & $ 18.1$ & E    & $0.16$ & $1.273$ & $1398$ & $0.838$ &               & \nl        
NGC 4482      & $-18.52$ & $13.34$ & $ 23.6$ & E/S0 & $0.41$ & $1.234$ & $1843$ &         &               & \nl        
NGC 4494      & $-20.84$ & $10.41$ & $ 17.8$ & E    & $0.05$ & $1.688$ & $1310$ & $0.810$ &               & $14.83$ \nl        
NGC 4503      & $-19.44$ & $11.79$ & $ 17.6$ & S0a  & $0.51$ & $1.560$ & $1360$ & $0.863$ &               & \nl        
NGC 4564      & $-19.15$ & $11.67$ & $ 14.6$ & E    & $0.45$ & $1.530$ & $1116$ & $0.863$ &               & \nl        
NGC 4589      & $-20.88$ & $11.33$ & $ 27.6$ & E    & $0.18$ & $1.563$ & $1945$ & $0.865$ &  $38 \pm 2$   & $15.21$\nl        
NGC 4621      & $-18.50$ & $10.41$ & $  6.0$ & E    & $0.24$ & $1.736$ & $ 431$ & $0.879$ &               & \nl        
NGC 4648      & $-19.09$ & $12.60$ & $ 21.8$ & E    & $0.26$ & $1.287$ & $1476$ & $0.829$ &               & \nl        
NGC 5017      & $-19.19$ & $13.25$ & $ 30.8$ & E    & $0.24$ & $1.255$ & $2534$ & $0.882$ &               & \nl        
NGC 5077      & $-20.63$ & $12.03$ & $ 34.0$ & E    & $0.20$ & $1.348$ & $2769$ & $0.942$ & $161 \pm 6$   & $12.02$ \nl        
NGC 5173      & $-19.52$ & $13.07$ & $ 32.8$ & E/S0 & $0.00$ & $1.043$ & $2419$ &         & $3.2 \pm 0.6$ & $16.37$ \nl        
NGC 5198      & $-20.24$ & $12.42$ & $ 34.1$ & E/S0 & $0.14$ & $1.334$ & $2514$ & $0.845$ & $4.0 \pm 0.6$ & \nl        
NGC 5283      & $-18.93$ & $14.13$ & $ 40.8$ & S0   & $0.10$ & $1.061$ & $3005$ &         & $13.4 \pm 0.4$ & \nl        
NGC 5308      & $-20.13$ & $11.99$ & $ 26.6$ & S0a  & $0.82$ & $1.563$ & $1877$ & $0.739$ &               & \nl        
NGC 5370      & $-19.45$ & $13.63$ & $ 41.3$ & S0   & $0.02$ & $1.145$ & $3053$ &         &               & \nl        
NGC 5557      & $-21.52$ & $11.62$ & $ 42.5$ & E    & $0.21$ & $1.405$ & $3210$ & $0.824$ &               & \nl        
NGC 5576      & $-20.00$ & $11.41$ & $ 19.1$ & E    & $0.23$ & $1.524$ & $1484$ & $0.807$ &               & \nl        
NGC 5796      & $-20.38$ & $12.43$ & $ 36.5$ & E    & $0.25$ & $1.434$ & $2961$ & $0.953$ & $110 \pm 4$   & \nl        
NGC 5812      & $-20.10$ & $11.86$ & $ 24.6$ & E    & $0.12$ & $1.420$ & $1961$ & $0.928$ &               & \nl        
NGC 5813      & $-20.85$ & $11.11$ & $ 24.6$ & E    & $0.24$ & $1.663$ & $1917$ & $0.892$ & $16 \pm 1$    & \nl        
NGC 5831      & $-19.59$ & $12.06$ & $ 21.4$ & E    & $0.13$ & $1.397$ & $1660$ & $0.874$ &               & \nl        
\tablebreak
NGC 5846      & $-21.10$ & $10.74$ & $ 23.3$ & E    & $0.06$ & $1.631$ & $1722$ & $0.912$ & $22 \pm 1$    & \nl        
NGC 5898      & $-20.30$ & $11.80$ & $ 26.3$ & E    & $0.07$ & $1.459$ & $2171$ & $0.929$ &               & \nl        
NGC 5903      & $-20.85$ & $11.58$ & $ 30.5$ & E    & $0.26$ & $1.530$ & $2512$ & $0.875$ & $32 \pm 2$    & \nl        
NGC 5982      & $-21.32$ & $11.65$ & $ 39.3$ & E    & $0.30$ & $1.516$ & $2876$ & $0.816$ &               & \nl        
NGC 6278      & $-19.70$ & $13.15$ & $ 37.1$ & S0   & $0.42$ & $1.345$ & $2791$ &         &               & \nl        
UGC 4551      & $-18.77$ & $13.09$ & $ 23.6$ & S0   & $0.67$ & $1.322$ & $1722$ & $0.813$ &               & \nl        
UGC 4587      & $-19.54$ & $13.49$ & $ 40.3$ & S0   & $0.43$ & $1.215$ & $3060$ &         &               & \nl        
UGC 5467      & $-18.87$ & $13.95$ & $ 36.7$ & S0   & $0.00$ & $1.022$ & $2894$ &         & $7.9 \pm 1.1$ & $13.40$ \nl 
UGC 6062      & $-18.82$ & $13.74$ & $ 32.7$ & S0   & $0.39$ & $1.137$ & $2607$ &         &               & \nl       
\enddata
\tablecomments{Global parameters   for the galaxies    in our sample.  
  Except for the data  in the last two columns,  these data are  taken
  from the Lyon/Meudon Extragalactic Database (LEDA):\\ 
Column (1): The name of the galaxy.\\  
Column (2): The absolute  $B$-band   magnitude, based on  the
  apparent  magnitudes and  distances listed in  columns  (3) and (4),
  respectively.\\
Column (3): The total apparent $B$-band magnitude corrected for
  galactic extinction and redshift effects (see Paturel \etal 1997).\\
Column (4): Distances  in  Mpc, as  derived from a  pure Hubble
  expansion   using   the  Virgo-centric   infall  corrected  velocity
  (column~[8]) and a Hubble constant of $H_0 = 80 \kmsmpc$.\\
Column (5): Morphological type, based on the code system of the
  Third  Reference Catalogue of  Bright Galaxies (de Vaucouleurs \etal
  1991). A  definition of this  classification can be found in Paturel
  \etal (1997).\\
Column (6): The  ellipticity of the $\mu_B = 25$ mag arcsec$^{-2}$
  isophote.\\
Column (7): The apparent diameter (in $^{10}{\rm log}$ of $0.1'$) of the
  galaxy according to the convention of the Second Reference Catalogue
  of Bright Galaxies (de Vaucouleurs \etal   1976).\\  
Column  (8): The radial  velocity  (in $\kms$)   of the galaxy
  corrected  for  Virgo-centric infall  (see  Paturel \etal 1997 for the
  infall model).\\
Column (9): Total B-V color, corrected for galactic extinction and
  redshift effects.\\
Column (10): Radio flux (in mJy) at 1.4 GHz from the NVSS (Condon
  \etal 1998).\\
Column (11): Far infrared magnitude, based on the IRAS 60
  $\mu$m and 100 $\mu$m fluxes (see equation~[\ref{mfir}]).\\
}
\end{deluxetable}


\begin{deluxetable}{lccccccc}
\tablewidth{0pt}
\tablecaption{Galaxy Morphology.\label{tab:isoparameter}}
\tablehead{\colhead{Name} & \colhead{Profile} &\colhead{Dust} & 
\colhead{Dust} & \colhead{IC} & \colhead{Nucl.} & 
\colhead{M.S.} & \colhead{$\langle \epsilon \rangle$} \\
\colhead{} & \colhead{} & \colhead{Morph.} & \colhead{Level} &
\colhead{} & \colhead{} & \colhead{} & 
\colhead{}  \\
\colhead{(1)} & \colhead{(2)} & \colhead{(3)} & \colhead{(4)} & 
\colhead{(5)} & \colhead{(6)} & \colhead{(7)} & \colhead{(8)}  \\
}
\startdata
ESO 378-20 & $\setminus$ &  &  & {\it d0d } &  & 2 & $0.34 \pm 0.02$ \nl
ESO 437-15 & ? & f & III & {\it xd } &  &  & $0.56 \pm 0.08$ \nl
ESO 443-39 & $\setminus$ &  &  & {\it 0d0d } &  & 2 & $0.25 \pm 0.04$ \nl
ESO 447-30 & $\setminus$ &  &  & {\it 0d0d } &  & 2 & $0.27 \pm 0.08$ \nl
ESO 507-27 & $\setminus$ &  &  & {\it dbd } & II &  & $0.49 \pm 0.07$ \nl
ESO 580-26 & ? & f & III & {\it x } &  &  & $0.47 \pm 0.11$ \nl
IC 875 & $\setminus$ &  &  & {\it d0 } &  &  & $0.55 \pm 0.05$ \nl
MCG 11-14-25A & ) &  &  & {\it 0 } &  &  & $0.11 \pm 0.02$ \nl
MCG 8-27-18 & $\setminus$ &  &  & {\it 0 } &  &  & $0.10 \pm 0.01$ \nl
NGC 2549 & $\setminus$ &  &  & {\it dbd } & II &  & $0.41 \pm 0.11$ \nl
NGC 2592 & $\setminus$ & d & 0.6 & {\it x0d } &  & 1 & $0.18 \pm 0.06$ \nl
NGC 2634 & $\setminus$ &  &  & {\it 0 } & I &  & $0.10 \pm 0.02$ \nl
NGC 2699 & $\setminus$ & d & 0.6 & {\it x0d0 } &  & 1 & $0.16 \pm 0.07$ \nl
NGC 2778 & $\setminus$ &  &  & {\it 0d0 } &  &  & $0.16 \pm 0.05$ \nl
NGC 2824 & ? & f & III & {\it x } &  &  & $0.41 \pm 0.20$ \nl
NGC 2872 & $\setminus$ & d & 0.8 & {\it x0 } &  &  & $0.23 \pm 0.01$ \nl
NGC 2902 & $\setminus$ & f & II & {\it x0 } &  &  & $0.05 \pm 0.02$ \nl
NGC 2950 & $\setminus$ &  &  & {\it d0d } &  & 2 & $0.29 \pm 0.04$ \nl
NGC 2986 & $\cap$ &  &  & {\it ?0 } &  &  & $0.18 \pm 0.01$ \nl
NGC 3065 & $\setminus$ & d & 0.4 & {\it x0d0 } &  &  & $0.10 \pm 0.06$ \nl
NGC 3078 & $\setminus$ & d & 1.2 & {\it xb } &  &  & $0.24 \pm 0.01$ \nl
NGC 3193 & ) &  &  & {\it 0 } &  &  & $0.20 \pm 0.03$ \nl
NGC 3226 & ? & f & III & {\it x } &  &  & $0.17 \pm 0.04$ \nl
NGC 3266 & $\setminus$ &  &  & {\it 0d0 } & II & 1 & $0.08 \pm 0.08$ \nl
NGC 3348 & $\cap$ &  &  & {\it 0 } &  &  & $0.07 \pm 0.01$ \nl
NGC 3377 & $\setminus$ & f & I & {\it d } &  &  & $0.53 \pm 0.01$ \nl
NGC 3414 & $\setminus$ & f & I & {\it x0 } &  &  & $0.26 \pm 0.02$ \nl
NGC 3595 & $\setminus$ &  &  & {\it dbd } &  & 1 & $0.34 \pm 0.04$ \nl
NGC 3613 & $\cap$ &  &  & {\it db0 } &  &  & $0.24 \pm 0.06$ \nl
\tablebreak
NGC 3640 & ) &  &  & {\it 0 } &  &  & $0.20 \pm 0.03$ \nl
NGC 4121 & $\setminus$ &  &  & {\it 0d0 } &  & 1 & $0.29 \pm 0.04$ \nl
NGC 4125 & ? & f & III & {\it x0 } &  &  & $0.34 \pm 0.12$ \nl
NGC 4128 & $\setminus$ &  &  & {\it dbd } &  &  & $0.39 \pm 0.02$ \nl
NGC 4168 & $\cap$ & f & II & {\it x0 } & II &  & $0.18 \pm 0.03$ \nl
NGC 4233 & ? & d & 12.3 & {\it x } &  &  & $0.30 \pm 0.18$ \nl
NGC 4291 & $\cap$ &  &  & {\it 0b } &  &  & $0.24 \pm 0.02$ \nl
NGC 4365 & $\cap$ &  &  & {\it ?d0 } &  &  & $0.29 \pm 0.03$ \nl
NGC 4474 & $\setminus$ &  &  & {\it dbd } & II &  & $0.20 \pm 0.09$ \nl
NGC 4478 & $\setminus$ &  &  & {\it db  } &  &  & $0.21 \pm 0.07$ \nl
NGC 4482 & $\setminus$ &  &  & {\it ?0 } & III &  & $0.27 \pm 0.08$ \nl
NGC 4494 & ? & d & 1.6 & {\it x0 } &  &  & $0.21 \pm 0.05$ \nl
NGC 4503 & $\setminus$ &  &  & {\it 0d0 } &  & 1 & $0.26 \pm 0.01$ \nl
NGC 4564 & $\setminus$ &  &  & {\it 0d } &  &  & $0.29 \pm 0.05$ \nl
NGC 4589 & ) & f & II & {\it xb } &  &  & $0.19 \pm 0.03$ \nl
NGC 4621 & $\setminus$ &  &  & {\it d } &  &  & $0.37 \pm 0.02$ \nl
NGC 4648 & $\setminus$ & d & 0.3 & {\it 0d0 } &  & 1 & $0.21 \pm 0.05$ \nl
NGC 5017 & $\setminus$ & d & 0.6 & {\it x0 } &  &  & $0.17 \pm 0.02$ \nl
NGC 5077 & ) & f & II & {\it xb } & II &  & $0.29 \pm 0.03$ \nl
NGC 5173 & ? & f & III & {\it x } &  &  & $0.14 \pm 0.02$ \nl
NGC 5198 & ) &  &  & {\it 0 } &  &  & $0.11 \pm 0.04$ \nl
NGC 5283 & ? & f & III & {\it x } &  &  & $0.13 \pm 0.06$ \nl
NGC 5308 & $\setminus$ &  &  & {\it d0d } &  &  & $0.47 \pm 0.02$ \nl
NGC 5370 & $\setminus$ &  &  & {\it 0d } &  &  & $0.18 \pm 0.11$ \nl
NGC 5557 & ) &  &  & {\it 0 } &  &  & $0.20 \pm 0.03$ \nl
NGC 5576 & $\setminus$ &  &  & {\it 0b0 } &  &  & $0.29 \pm 0.03$ \nl
NGC 5796 & $\setminus$ & f & I & {\it 0 } &  &  & $0.14 \pm 0.01$ \nl
NGC 5812 & $\setminus$ & d & 0.4 & {\it x0 } &  &  & $0.02 \pm 0.01$ \nl
NGC 5813 & $\cap$ & f & II & {\it x0 } &  &  & $0.10 \pm 0.01$ \nl
NGC 5831 & $\setminus$ &  &  & {\it 0 } &  &  & $0.30 \pm 0.03$ \nl
\tablebreak
NGC 5846 & ? & f & III & {\it x0 } &  &  & $0.08 \pm 0.03$ \nl
NGC 5898 & $\setminus$ & f & I & {\it 0 } &  &  & $0.16 \pm 0.06$ \nl
NGC 5903 & $\cap$ & f & I & {\it ?b } &  &  & $0.22 \pm 0.03$ \nl
NGC 5982 & $\cap$ &  &  & {\it 0b } &  &  & $0.31 \pm 0.07$ \nl
NGC 6278 & $\setminus$ &  &  & {\it b0d } &  & 1 & $0.22 \pm 0.09$ \nl
UGC 4551 & $\setminus$ &  &  & {\it 0d } & I & 1 & $0.15 \pm 0.04$ \nl
UGC 4587 & $\setminus$ & f & I & {\it 0 } &  &  & $0.28 \pm 0.07$ \nl
UGC 6062 & $\setminus$ &  &  & {\it 0d0d } &  & 2 & $0.30 \pm 0.04$ \nl
\enddata
\tablecomments{Isophotal parameters  for  all the  67 galaxies in  our
  sample.  Column (2)  indicates the classification where $\setminus$,
  $)$, and $\cap$ indicate power-law,  intermediate and core galaxies. 
  A question  mark indicates  that  no  meaningful luminosity  profile
  could be  determined (either due  to dust   or nucleation)  and  the
  galaxy classification is  therefore unknown in  these cases.  Column
  (3) gives the dust morphology where `d' stands for dust disk and `f'
  stands  for dust filaments.  Column (4)  indicates either the amount
  of filamentary  dust ranging  from I  (small) to  III (large) or  the
  diameter of  the  dust disk in  arcsec,  if applicable.  Column  (5)
  gives the isophotal code (IC) of  the galaxy going from $r=0.2''$ to
  the outer region where $d$,  $b$, $0$, $x$,  and $?$ indicate disky,
  boxy,  neither   disky  nor  boxy,  undetermined  due to  dust,  and
  undetermined due to small surface   brightness gradient. Column  (6)
  indicates  the degree of nucleation,   ranging from I  (weak) to III
  (strong).  Column (7)     indicates  if and  how   many   misaligned
  structures   (M.S.)    are    detected   in     the    galaxy   (see
  \S~\ref{sec:misdisk}).     Column (8),  finally,    lists the   mean
  ellipticity   (determined over  the radial interval    $1.0'' < r  <
  10.0''$) and its standard deviation.}
\end{deluxetable}


\begin{deluxetable}{lccccccc}
\tablecaption{Parameters of major axis luminosity profiles.\label{tab:NukerMajor}}
\tablehead{
\colhead{Name} & \colhead{Profile} & \colhead{$I_b$} & \colhead{$\alpha$} &
\colhead{$\beta$} & \colhead{$\gamma$} & \colhead{$r_b$} & 
\colhead{$\gamma '$}  \\
\colhead{(1)} & \colhead{(2)} & \colhead{(3)} & \colhead{(4)} & 
\colhead{(5)} & \colhead{(6)} & \colhead{(7)} & \colhead{(8)} \\
}
\startdata
ESO 378-20 & $\setminus$ & 17.61 & 0.43 & 2.00 & 0.86 & 2.41 & 1.09 \nl
ESO 437-15 & ? &  &  &  &  &  &  \nl
ESO 443-39 & $\setminus$ & 17.32 & 1.36 & 1.31 & 0.75 & 1.30 & 0.77 \nl
ESO 447-30 & $\setminus$ & 16.73 & 2.70 & 1.72 & 0.84 & 1.50 & 0.84 \nl
ESO 507-27 & $\setminus$ & 17.96 & 0.59 & 1.58 & 0.70 & 3.89 & 0.79 \nl
ESO 580-26 & ? &  &  &  &  &  &  \nl
IC 875 & $\setminus$ & 17.65 & 0.85 & 1.81 & 1.07 & 1.87 & 1.12 \nl
MCG 11-14-25A & ) & 16.59 & 0.69 & 2.13 & 0.00 & 0.56 & 0.49 \nl
MCG 8-27-18 & $\setminus$ & 18.61 & 0.80 & 1.93 & 0.79 & 1.95 & 0.89 \nl
NGC 2549 & $\setminus$ & 16.96 & 1.75 & 1.71 & 0.67 & 3.70 & 0.67 \nl
NGC 2592 & $\setminus$ & 16.61 & 3.31 & 1.60 & 0.92 & 1.37 & 0.92 \nl
NGC 2634 & $\setminus$ & 17.32 & 2.83 & 1.57 & 0.81 & 1.82 & 0.81 \nl
NGC 2699 & $\setminus$ & 17.63 & 1.66 & 1.89 & 0.84 & 2.66 & 0.85 \nl
NGC 2778 & $\setminus$ & 17.08 & 0.93 & 1.55 & 0.94 & 1.33 & 0.99 \nl
NGC 2824 & ? &  &  &  &  &  &  \nl
NGC 2872 & $\setminus$ & 18.11 & 2.55 & 1.66 & 1.01 & 4.27 & 1.01 \nl
NGC 2902 & $\setminus$ & 16.34 & 1.93 & 1.57 & 0.49 & 0.88 & 0.50 \nl
NGC 2950 & $\setminus$ & 16.16 & 2.40 & 1.81 & 0.82 & 2.43 & 0.82 \nl
NGC 2986 & $\cap$ & 16.09 & 1.77 & 1.50 & 0.18 & 1.24 & 0.20 \nl
NGC 3065 & $\setminus$ & 17.97 & 0.99 & 2.28 & 0.79 & 3.14 & 0.84 \nl
NGC 3078 & $\setminus$ & 17.19 & 2.35 & 1.60 & 0.95 & 3.43 & 0.95 \nl
NGC 3193 & ) & 15.46 & 0.59 & 1.89 & 0.01 & 0.81 & 0.44 \nl
NGC 3226 & ? &  &  &  &  &  &  \nl
NGC 3266 & $\setminus$ & 17.78 & 1.34 & 2.06 & 0.64 & 2.25 & 0.66 \nl
NGC 3348 & $\cap$ & 16.00 & 1.17 & 1.53 & 0.09 & 0.99 & 0.18 \nl
NGC 3377 & $\setminus$ & 15.37 & 3.15 & 1.35 & 1.12 & 1.42 & 1.12 \nl
NGC 3414 & $\setminus$ & 16.46 & 1.40 & 1.45 & 0.83 & 1.72 & 0.84 \nl
NGC 3595 & $\setminus$ & 17.43 & 2.33 & 1.52 & 0.75 & 2.30 & 0.76 \nl
NGC 3613 & $\cap$ & 15.11 & 1.53 & 1.06 & 0.04 & 0.34 & 0.17 \nl
\tablebreak
NGC 3640 & ) & 15.24 & 0.87 & 1.16 & 0.15 & 0.44 & 0.37 \nl
NGC 4121 & $\setminus$ & 19.69 & 1.51 & 3.65 & 0.85 & 5.60 & 0.85 \nl
NGC 4125 & ? &  &  &  &  &  &  \nl
NGC 4128 & $\setminus$ & 16.31 & 1.13 & 1.69 & 0.71 & 1.59 & 0.75 \nl
NGC 4168 & $\cap$ & 17.45 & 1.43 & 1.39 & 0.17 & 2.02 & 0.19 \nl
NGC 4233 & ? &  &  &  &  &  &  \nl
NGC 4291 & $\cap$ & 15.06 & 1.35 & 1.62 & 0.00 & 0.60 & 0.13 \nl
NGC 4365 & $\cap$ & 16.08 & 2.29 & 1.28 & 0.13 & 1.82 & 0.13 \nl
NGC 4474 & $\setminus$ & 17.97 & 2.28 & 1.65 & 0.72 & 4.55 & 0.72 \nl
NGC 4478 & $\setminus$ & 17.57 & 0.93 & 1.73 & 0.57 & 4.16 & 0.61 \nl
NGC 4482 & $\setminus$ & 19.63 & 3.43 & 1.01 & 0.49 & 4.04 & 0.49 \nl
NGC 4494 & ? & 15.21 &  & 1.25 &  &  &  \nl
NGC 4503 & $\setminus$ & 16.53 & 1.77 & 1.30 & 0.64 & 1.65 & 0.65 \nl
NGC 4564 & $\setminus$ & 15.89 & 1.43 & 1.27 & 0.80 & 1.28 & 0.81 \nl
NGC 4589 & ) & 15.64 & 0.82 & 1.44 & 0.26 & 0.71 & 0.46 \nl
NGC 4621 & $\setminus$ & 16.57 & 2.53 & 1.40 & 0.96 & 3.75 & 0.96 \nl
NGC 4648 & $\setminus$ & 15.87 & 3.73 & 1.54 & 0.92 & 1.02 & 0.92 \nl
NGC 5017 & $\setminus$ & 16.92 & 2.97 & 1.59 & 1.12 & 1.48 & 1.12 \nl
NGC 5077 & ) & 16.49 & 1.09 & 1.67 & 0.23 & 1.61 & 0.30 \nl
NGC 5173 & ? &  &  &  &  &  &  \nl
NGC 5198 & ) & 14.79 & 2.61 & 1.13 & 0.23 & 0.16 & 0.42 \nl
NGC 5283 & ? &  &  &  &  &  &  \nl
NGC 5308 & $\setminus$ & 15.55 & 0.39 & 1.27 & 0.82 & 0.83 & 0.96 \nl
NGC 5370 & $\setminus$ & 17.84 & 0.78 & 1.50 & 0.62 & 1.63 & 0.71 \nl
NGC 5557 & ) & 16.18 & 0.80 & 1.77 & 0.14 & 1.21 & 0.33 \nl
NGC 5576 & $\setminus$ & 15.79 & 0.79 & 1.73 & 0.36 & 1.41 & 0.51 \nl
NGC 5796 & $\setminus$ & 16.19 & 0.83 & 1.67 & 0.41 & 1.06 & 0.56 \nl
NGC 5812 & $\setminus$ & 16.61 & 1.25 & 1.67 & 0.59 & 1.84 & 0.62 \nl
NGC 5813 & $\cap$ & 15.77 & 1.90 & 1.33 & 0.00 & 0.73 & 0.03 \nl
NGC 5831 & $\setminus$ & 16.90 & 0.47 & 1.84 & 0.33 & 1.78 & 0.64 \nl
\tablebreak
NGC 5846 & ? &  &  &  &  &  &  \nl
NGC 5898 & $\setminus$ & 16.32 & 1.23 & 1.57 & 0.41 & 1.24 & 0.46 \nl
NGC 5903 & $\cap$ & 16.84 & 1.80 & 1.48 & 0.13 & 1.59 & 0.14 \nl
NGC 5982 & $\cap$ & 15.60 & 0.99 & 1.52 & 0.00 & 0.74 & 0.18 \nl
NGC 6278 & $\setminus$ & 15.69 & 0.76 & 1.62 & 0.55 & 0.60 & 0.77 \nl
UGC 4551 & $\setminus$ & 17.06 & 2.19 & 2.16 & 0.51 & 2.26 & 0.51 \nl
UGC 4587 & $\setminus$ & 15.48 & 1.04 & 1.24 & 0.72 & 0.19 & 0.90 \nl
UGC 6062 & $\setminus$ & 18.35 & 0.90 & 1.81 & 0.80 & 2.75 & 0.85 \nl
\enddata
\tablecomments{Parameterizations    of    the  major  axis  luminosity
  profiles.  Columns  (2) indicates  the classification of  the galaxy
  (copied   from  column~(2)   in    Table~\ref{tab:isoparameter}).    
  Columns~(3)  - (7)  list  the best fit   parameters of the Nuker-law
  (equation~[\ref{nukerlaw}]).   Here the break  radius, $r_b$,  is in
  arcsec and $I_b$ in mag arcsec$^{-2}$.  Column~(8)  lists the
  value of  $\gamma'$    (see
  \S~\ref{sec:lumprof} for definition).}
\end{deluxetable}


\begin{deluxetable}{lccccccc}
\tablecaption{Parameters of minor axis luminosity profiles. \label{tab:NukerMinor}}
\tablehead{
\colhead{Name} & \colhead{Profile} & \colhead{$I_b$} & \colhead{$\alpha$} &
\colhead{$\beta$} & \colhead{$\gamma$} & \colhead{$r_b$} &
\colhead{$\gamma '$}\\
\colhead{(1)} & \colhead{(2)} & \colhead{(3)} & \colhead{(4)} & 
\colhead{(5)} & \colhead{(6)} & \colhead{(7)} & \colhead{(8)}  \\
}
\startdata
ESO 378-20 & $\setminus$ & 17.96 & 0.39 & 2.06 & 0.87 & 1.98 & 1.15 \nl
ESO 437-15 & ? &  &  &  &  &  &  \nl
ESO 443-39 & $\setminus$ & 17.52 & 1.53 & 1.63 & 0.75 & 1.19 & 0.77 \nl
ESO 447-30 & $\setminus$ & 16.88 & 1.31 & 1.60 & 0.83 & 1.19 & 0.86 \nl
ESO 507-27 & $\setminus$ & 17.71 & 1.56 & 2.25 & 0.67 & 1.66 & 0.69 \nl
ESO 580-26 & ? &  &  &  &  &  &  \nl
IC 875 & $\setminus$ & 18.93 & 0.41 & 1.78 & 1.04 & 2.14 & 1.21 \nl
MCG 11-14-25A & ) & 16.69 & 0.70 & 2.21 & 0.00 & 0.56 & 0.51 \nl
MCG 8-27-18 & $\setminus$ & 18.39 & 0.82 & 1.87 & 0.79 & 1.53 & 0.89 \nl
NGC 2549 & $\setminus$ & 15.88 & 3.49 & 1.55 & 0.63 & 0.94 & 0.63 \nl
NGC 2592 & $\setminus$ & 17.40 & 2.40 & 1.86 & 1.06 & 1.92 & 1.06 \nl
NGC 2634 & $\setminus$ & 17.33 & 2.56 & 1.64 & 0.79 & 1.66 & 0.80 \nl
NGC 2699 & $\setminus$ & 17.87 & 1.06 & 1.84 & 0.95 & 2.40 & 0.98 \nl
NGC 2778 & $\setminus$ & 16.56 & 1.49 & 1.53 & 0.96 & 0.77 & 0.98 \nl
NGC 2824 & ? &  &  &  &  &  &  \nl
NGC 2872 & $\setminus$ & 17.84 & 3.12 & 1.58 & 1.00 & 2.76 & 1.00 \nl
NGC 2902 & $\setminus$ & 16.27 & 2.14 & 1.50 & 0.49 & 0.77 & 0.51 \nl
NGC 2950 & $\setminus$ & 16.16 & 1.66 & 1.78 & 0.84 & 1.65 & 0.84 \nl
NGC 2986 & $\cap$ & 16.05 & 1.86 & 1.45 & 0.19 & 0.97 & 0.20 \nl
NGC 3065 & $\setminus$ & 18.04 & 0.86 & 2.17 & 0.83 & 2.97 & 0.89 \nl
NGC 3078 & $\setminus$ & 17.49 & 3.42 & 1.66 & 1.06 & 3.26 & 1.06 \nl
NGC 3193 & ) & 14.21 & 1.18 & 1.18 & 0.00 & 0.14 & 0.48 \nl
NGC 3226 & ? &  &  &  &  &  &  \nl
NGC 3266 & $\setminus$ & 17.28 & 2.39 & 1.91 & 0.67 & 1.46 & 0.67 \nl
NGC 3348 & $\cap$ & 16.06 & 1.08 & 1.61 & 0.06 & 0.98 & 0.18 \nl
NGC 3377 & $\setminus$ & 15.91 &  & 1.29 &  &  &  \nl
NGC 3414 & $\setminus$ & 16.29 & 1.13 & 1.39 & 0.85 & 1.10 & 0.89 \nl
NGC 3595 & $\setminus$ & 17.26 & 2.43 & 1.75 & 0.77 & 1.34 & 0.77 \nl
NGC 3613 & $\cap$ & 15.73 & 1.23 & 1.46 & 0.13 & 0.65 & 0.25 \nl
\tablebreak
NGC 3640 & ) & 15.41 & 0.86 & 1.28 & 0.17 & 0.48 & 0.40 \nl
NGC 4121 & $\setminus$ & 19.32 & 1.28 & 2.90 & 0.85 & 3.48 & 0.87 \nl
NGC 4125 & ? &  &  &  &  &  &  \nl
NGC 4128 & $\setminus$ & 17.30 & 0.83 & 2.22 & 0.67 & 1.98 & 0.79 \nl
NGC 4168 & $\cap$ & 17.48 & 1.43 & 1.30 & 0.21 & 1.64 & 0.23 \nl
NGC 4233 & ? &  &  &  &  &  &  \nl
NGC 4291 & $\cap$ & 15.07 & 1.36 & 1.65 & 0.00 & 0.47 & 0.18 \nl
NGC 4365 & $\cap$ & 16.25 & 1.54 & 1.39 & 0.11 & 1.55 & 0.13 \nl
NGC 4474 & $\setminus$ & 17.43 & 3.83 & 1.95 & 0.58 & 2.38 & 0.58 \nl
NGC 4478 & $\setminus$ & 17.63 & 0.87 & 1.61 & 0.54 & 3.31 & 0.59 \nl
NGC 4482 & $\setminus$ & 19.92 & 2.17 & 1.50 & 0.50 & 4.02 & 0.50 \nl
NGC 4494 & ? & 15.58 &  & 1.18 &  &  &  \nl
NGC 4503 & $\setminus$ & 16.89 & 1.21 & 1.60 & 0.63 & 1.70 & 0.66 \nl
NGC 4564 & $\setminus$ & 16.90 & 1.13 & 1.93 & 0.78 & 2.10 & 0.82 \nl
NGC 4589 & ) & 15.64 & 0.71 & 1.45 & 0.18 & 0.54 & 0.47 \nl
NGC 4621 & $\setminus$ & 17.52 & 0.69 & 1.49 & 0.96 & 4.81 & 1.00 \nl
NGC 4648 & $\setminus$ & 16.50 & 3.16 & 1.75 & 1.03 & 1.29 & 1.03 \nl
NGC 5017 & $\setminus$ & 17.75 & 1.81 & 1.68 & 1.19 & 2.12 & 1.19 \nl
NGC 5077 & ) & 16.48 & 1.15 & 1.74 & 0.25 & 1.19 & 0.33 \nl
NGC 5173 & ? &  &  &  &  &  &  \nl
NGC 5198 & ) & 14.81 & 2.04 & 1.15 & 0.21 & 0.14 & 0.52 \nl
NGC 5283 & ? &  &  &  &  &  &  \nl
NGC 5308 & $\setminus$ & 18.39 & 0.85 & 2.18 & 0.86 & 3.95 & 0.92 \nl
NGC 5370 & $\setminus$ & 18.61 & 1.26 & 2.51 & 0.64 & 2.40 & 0.68 \nl
NGC 5557 & ) & 15.74 & 1.11 & 1.54 & 0.18 & 0.67 & 0.33 \nl
NGC 5576 & $\setminus$ & 15.57 & 0.91 & 1.76 & 0.36 & 0.89 & 0.53 \nl
NGC 5796 & $\setminus$ & 16.35 & 0.82 & 1.71 & 0.40 & 1.05 & 0.57 \nl
NGC 5812 & $\setminus$ & 16.69 & 1.22 & 1.73 & 0.58 & 1.91 & 0.62 \nl
NGC 5813 & $\cap$ & 15.88 & 1.61 & 1.41 & 0.00 & 0.74 & 0.05 \nl
NGC 5831 & $\setminus$ & 15.16 & 0.99 & 1.34 & 0.39 & 0.27 & 0.65 \nl
\tablebreak
NGC 5846 & ? &  &  &  &  &  &  \nl
NGC 5898 & $\setminus$ & 16.63 & 0.90 & 1.61 & 0.38 & 1.35 & 0.49 \nl
NGC 5903 & $\cap$ & 16.66 & 2.30 & 1.35 & 0.12 & 0.98 & 0.13 \nl
NGC 5982 & $\cap$ & 15.26 & 1.91 & 1.42 & 0.05 & 0.46 & 0.12 \nl
NGC 6278 & $\setminus$ & 16.31 & 1.11 & 1.96 & 0.65 & 0.86 & 0.76 \nl
UGC 4551 & $\setminus$ & 17.39 & 1.95 & 2.57 & 0.49 & 2.48 & 0.49 \nl
UGC 4587 & $\setminus$ & 15.97 & 0.72 & 1.47 & 0.63 & 0.23 & 0.92 \nl
UGC 6062 & $\setminus$ & 18.46 & 1.05 & 2.05 & 0.80 & 2.16 & 0.85 \nl
\enddata
\tablecomments{Same as Table~\ref{tab:NukerMajor},  except now for the
  minor axis luminosity profiles.}
\end{deluxetable}


\begin{deluxetable}{l|rrrr|r}
\tablecaption{\label{tab:lumisocorr}}
\tablehead{
\colhead{Profile} & 
\colhead{d} & \colhead{db} & \colhead{b} &
\colhead{0} & \colhead{Total} \\
}
\startdata
Power-law    &  20 & 7 & 2 & 12 & 41 \\
Intermediate &   0 & 0 & 2 &  5 &  7 \\
Core         &   1 & 1 & 3 &  4 &  9 \\
\hline
Total        &  21 & 8 & 7 & 21 & 57 \\
\enddata
\tablecomments{Correlation   of luminosity   profiles   with isophotal
  structure.  For    each class  of luminosity    profiles (power-law,
  intermediate,  and core) we indicate  the number of galaxies (in the
  unperturbed sample)   with  a common isophotal  characteristic.   We
  distinguish between  disky   `d',  boxy `b', disky-boxy  `db',   and
  regular `0' (see \S~\ref{sec:isoshapes} for definitions).  Note that
  most  disky and disky-boxy  galaxies are power-law galaxies, whereas
  most boxy  galaxies  are classified  as either core  or intermediate
  galaxies}
\end{deluxetable}

\clearpage


\end{document}